\documentclass[12pt]{article}
\usepackage{epsfig,amsfonts,amssymb}
\usepackage{hyperref}
\usepackage{cite}
\input epsf.sty
\usepackage{cite}

\def\ZZZ{{\hbox{ Z\kern-1.6mm Z}}}
\def\RRR{{\hbox{ R\kern-2.4mm R}}}
\def\CCC{{\hbox{ C\kern-2.0mm C}}}
\def\zzz{{\hbox{z\kern-1mm z}}}

\newcommand{\nn}{\nonumber \\}

\newcommand{\qeq}{{\hbox{=\kern-2.3mm ? \kern.5mm }}}
\renewcommand{\qeq}{=}

\newcommand{\HH}{{\cal H}}

\newcommand{\wt}{\widetilde}
\newcommand{\wh}{\widehat}

\newcommand{\be}{\begin{equation}}
\newcommand{\ee}{\end{equation}}
\newcommand{\ben}{\begin{eqnarray}\displaystyle}
\newcommand{\een}{\end{eqnarray}}

\newcommand{\refb}[1]{(\ref{#1})}
\newcommand{\p}{\partial}
\newcommand{\sectiono}[1]{\section{#1}\setcounter{equation}{0}}

\newcommand\bOm{\bar\Omega}

\def\one{{\hbox{ 1\kern-.8mm l}}}
\def\zero{{\hbox{ 0\kern-1.5mm 0}}}

\begin{document}

\begin{flushright}
arXiv:1112.2515
\end{flushright}

\baselineskip 24pt

\begin{center}
{\Large \bf  Equivalence of Three Wall Crossing Formul\ae}

\end{center}

\vskip .6cm
\medskip

\vspace*{4.0ex}

\baselineskip=18pt

\centerline{\large \rm Ashoke Sen}

\vspace*{4.0ex}

\centerline{\large \it Harish-Chandra Research Institute}
\centerline{\large \it  Chhatnag Road, Jhusi,
Allahabad 211019, India}

\vspace*{1.0ex}
\centerline{\small E-mail:  sen@mri.ernet.in}

\vspace*{5.0ex}

\centerline{\bf Abstract} \bigskip

The wall crossing formula of Kontsevich and Soibelman gives an
implicit relation between the BPS indices on two sides of the wall of
marginal stability by equating two symplectomorphisms constructed from
the indices on two sides of the wall. The wall
crossing formul\ae\ of Manschot, Pioline and the author 
give two apparently different explicit expressions
for the BPS index on one side of the wall in terms of the BPS
indices on the other
side. 
We prove the equivalence of  all the three formul\ae.

\vfill \eject

\baselineskip=18pt

\tableofcontents

\sectiono{Introduction} \label{s1}

The central objects in a wall crossing formula are a
BPS index in some Hilbert space $\HH$, and a moduli space over which
the Hilbert space could vary. The BPS index remains constant over most of
the moduli space but could jump across certain codimension one subspaces
known as the walls of marginal stability. When this happens, 
wall crossing formula gives a relation between the BPS indices
on two sides of the wall\cite{Cecotti:1992rm,Seiberg:1994rs,
Seiberg:1994aj,Ferrari:1996sv,Bilal:1996sk, Douglas:2000gi,
Denef:2000nb,Denef:2002ru,Bates:2003vx,Denef:2007vg,ks,Kontsevich:2009xt,
1006.2706,MR2357325,Joyce:2008pc,Joyce:2009xv,Diaconescu:2007bf,
Gaiotto:2008cd,Alexandrov:2008gh,deBoer:2008zn,
Jafferis:2008uf,Cecotti:2009uf,Stoppa:2009,Toda:2009,Chuang:2010wx,
Cecotti:2010fi,Gaiotto:2010be,Andriyash:2010qv,
Manschot:2009ia,Manschot:2010qz,1102.1729,1103.0261,
Manschot:2011xc,1107.0723,1109.4941,1110.0466,1110.1619,
1111.6979,1112.2174}. 

In situations relevant to supersymmetric string theory / gauge theory 
the Hilbert space that is of relevance is the space of quantum states
carrying some fixed set of gauge charges, collectively denoted by a 
vector $\gamma$. The moduli space is parametrized by the asymptotic
values of certain scalar fields of the theory.
If the theory also contains some global conserved $U(1)$ charge $Q$, 
then we can 
define a refined index $\Omega_{\rm ref}(\gamma, y)$ which 
computes the index weighted by $y^Q$ for some continuous variable $y$.
 Under certain circumstances this refined index also
remains constant over most of the moduli space
and jumps only across the
walls of marginal stability. For example
in supersymmetric gauge theories we can define such an index by
taking $Q$ to be an appropriate
linear combination of the angular momentum and R-symmetry 
generator\cite{Gaiotto:2010be}.
In string theory, there are no global R-symmetry charges,
but we can define a refined index
by taking $Q$ to be one of the angular momentum 
generators\cite{Dimofte:2009bv,Dimofte:2009tm}. 
Such an index is
not protected under a change in the string coupling, \i.e.\ it can jump
even without crossing a wall of marginal stability, but we could nevertheless
study its  jump across the walls of
marginal stability keeping the string coupling fixed at some small value.
A refined wall crossing formula is a relation between the refined indices on two
sides of the wall of marginal stability. This is more general than ordinary (also
referred to as `numerical') wall
crossing formula, since by setting $y=1$ in the former we recover the latter.

The known wall crossing formul\ae\ take simpler form in terms of
the `rational refined index' defined as\cite{MR2357325,Joyce:2008pc,Joyce:2009xv,
Manschot:2010xp,Manschot:2010nc,
Nishinaka:2010fh}
\be \label{esecond}
\bar \Omega_{\rm ref}(\gamma,y) 
\equiv \sum_{m|\gamma} 
\frac{y - y^{-1}}{m\, (y^m - y^{-m})}\, 
\Omega_{\rm ref}(\gamma/m, y^m) \, .
\ee
In the $y\to 1$ limit this gives $\bar\Omega(\gamma) =\sum_{m|\gamma} 
m^{-2}\Omega(\gamma/m)$.
We shall denote by $\Omega_{\rm ref}^\pm(\gamma, y)$ the refined
indices on two sides of a wall of marginal stability, and by
$\bar \Omega_{\rm ref}^\pm(\gamma,y)$ the corresponding rational
refined indices. A wall crossing formula corresponds to a relation
between $\Omega^+_{\rm ref}$ and $\Omega^-_{\rm ref}$, or equivalently
between $\bar\Omega^+_{\rm ref}$ and $\bar\Omega^-_{\rm ref}$.
We shall work with the indices $\bar\Omega^\pm_{\rm ref}$ but if needed we can
invert' \refb{esecond} to calculate $\Omega^\pm_{\rm ref}$ in terms of
$\bar\Omega^\pm_{\rm ref}$\cite{Manschot:2010qz,Manschot:2011xc}.

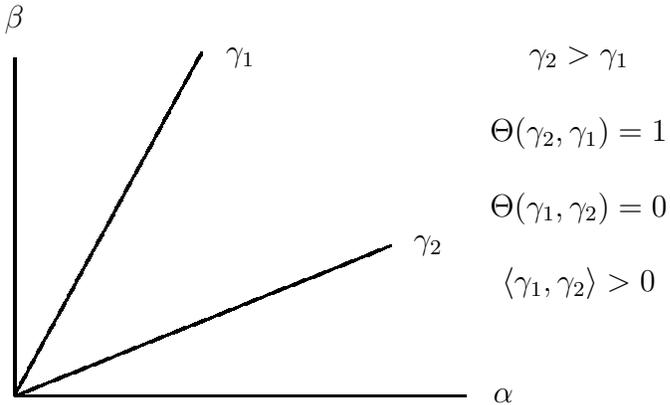
\begin{figure}
%
\ifx\JPicScale\undefined\def\JPicScale{1}\fi
\unitlength \JPicScale mm
\begin{picture}(135,90)(0,0)
\linethickness{0.3mm}
\put(60,40){\line(0,1){45}}
\linethickness{0.3mm}
\linethickness{0.3mm}
\put(60,40){\line(1,0){60}}
\linethickness{0.3mm}
\multiput(60,40)(0.12,0.22){208}{\line(0,1){0.22}}
\linethickness{0.3mm}
\multiput(60,40)(0.3,0.12){167}{\line(1,0){0.3}}
\put(85,85){\makebox(0,0)[cc]{}}

\put(90,85){\makebox(0,0)[cc]{$\gamma_1$}}

\put(115,60){\makebox(0,0)[cc]{$\gamma_2$}}

\put(60,90){\makebox(0,0)[cc]{$\beta$}}

\put(125,40){\makebox(0,0)[cc]{$\alpha$}}

\put(135,85){\makebox(0,0)[cc]{$\gamma_2>\gamma_1$}}

\put(135,75){\makebox(0,0)[cc]{$\Theta(\gamma_2,\gamma_1)=1$}}

\put(135,65){\makebox(0,0)[cc]{$\Theta(\gamma_1,\gamma_2)=0$}}

\put(135,55){\makebox(0,0)[cc]{$\langle\gamma_1,\gamma_2\rangle>0$}}

\end{picture}

\vskip -1.3in
\caption{Figure illustrating the definition of $\gamma_2>\gamma_1$ and
$\Theta(\gamma_2,\gamma_1)$.}
\label{f1}
\end{figure}

The charge $\gamma$ is a member of some charge lattice
equipped with a symplectic inner product.
Given a pair of vectors
$\gamma_1, \gamma_2$ on the charge lattice,
we denote by $\langle\gamma_1, \gamma_2\rangle$ the
symplectic inner product between $\gamma_1$ and $\gamma_2$.
This inner product is anti-symmetric under the exchange of 
$\gamma_1$ and $\gamma_2$ and is linear in $\gamma_1$, $\gamma_2$.
Typically the charge lattice has dimension
$d$ for some even integer $d$, but for a given wall of marginal stability
the relevant charge 
vectors for which the index jumps across
the wall
are of the form $r \alpha +s\beta$, where $\alpha$, $\beta$  
are two vectors
whose central charges align at the wall and
$r$ and
$s$ are two non-negative rational numbers. 
We denote by $\Lambda$ the set
of all such non-zero charge vectors in the lattice.
Without any loss of generality we can choose a convention in which
 $\langle\beta, \alpha\rangle>0$ 
and represent $r\alpha + s\beta\in\Lambda$ by a vector $(r,s)$ in a two
dimensional plane. In this convention all the elements of $\Lambda$ are
represented as vectors in the first quadrant of this two dimensional plane,
and given $\gamma_1,\gamma_2\in\Lambda$, 
$\gamma_1$ and $\gamma_2$ are arranged in a clockwise (anti-clockwise)
order if $\langle \gamma_1, \gamma_2\rangle>0$ 
($\langle \gamma_1, \gamma_2\rangle<0$) (see Fig.~\ref{f1}). 
We denote
by $Z_\gamma$ the central charge for any vector $\gamma\in\Lambda$ --
a function of the moduli and a linear function of $\gamma$ such that the mass
of a BPS state of charge $\gamma$ is given by $|Z_\gamma|$ --
and
choose our convention such that
\be \label{eside}
\langle \gamma_1 , \gamma_2\rangle \, {\rm Im} \, (Z_{\gamma_1} \bar
Z_{\gamma_2}) < 0\, , \qquad \forall \quad 
\gamma_1, \gamma_2\in\Lambda\, ,
\ee
on the side of the wall in which we label the index by $\Omega^+_{\rm ref}$.
On the other side of the wall $\langle \gamma_1 , \gamma_2\rangle \, 
{\rm Im} \, (Z_{\gamma_1} \bar
Z_{\gamma_2}) > 0$ and the refined index is denoted by
$\Omega^-_{\rm ref}$.
Then the wall crossing formula, written in the notation of
\cite{Manschot:2010qz}, takes the form:
\be \label{efirst3}
\bar\Omega_{\rm ref}^-(\gamma,y) =
\sum_{n\geq 1 }\, 
  \sum_{\hbox{\small \tiny unordered set}\, \alpha_1,\dots, \alpha_n\in\Lambda \atop
\alpha_1+\dots +\alpha_n=\gamma}\, 
\frac{g_{\rm ref}(\{\alpha_i\},y)}{|{\rm Aut}(\{\alpha_i\})|}\, 
 \prod\nolimits_{i=1}^n \bOm_{\rm ref}^+(\alpha_i,y) \ , \ee
  where $g_{\rm ref}(\{\alpha_i\},y)$ is a function to be specified
 later, and
 $|{\rm Aut}(\{\alpha_i\})|$ is a symmetry factor defined as follows.
 If the set  
 $\{\alpha_i\}$ consists of $m_1$ copies of $\beta_1$,
 $m_2$ copies of $\beta_2$ etc. then
 $|{\rm Aut}(\{\alpha_i\})|=\prod_k m_k!$. 
 The sum in  \refb{efirst3} runs over all possible unordered decompositions
 of the vector $\gamma$ into the vectors $\alpha_1,\cdots \alpha_n\in\Lambda$.
 However this can also be rewritten as a sum over the ordered decomposition
 of the 
 vector $\gamma$ into the vectors $\alpha_1,\cdots \alpha_n\in\Lambda$
  as follows:
 \be \label{efirst3new}
\bar\Omega_{\rm ref}^-(\gamma,y) = \sum_{n\geq 1 }\, {1\over n!}\, 
  \sum_{\hbox{\small ordered decomposition} \,
\alpha_1+\dots +\alpha_n=\gamma}\, 
g_{\rm ref}(\{\alpha_i\},y)\, 
 \prod\nolimits_{i=1}^n \bOm_{\rm ref}^+(\alpha_i,y) \ .
 \ee
  For a single argument
 $g_{\rm ref}(\alpha; y)$ is taken to be 1, so that the $n=1$ term on the
 right hand side of \refb{efirst3} just gives
$ \bar\Omega^+_{\rm ref}
(\gamma,y)$.
 The wall crossing formula
 for rational numerical
 index can be found by taking the $y\to 1$ limit of the above
 formula.
 
 One of the results discussed in \cite{Manschot:2010qz} is that once we
 use the index $\bar\Omega$ instead of $\Omega$, the effect of having
 two or more identical $\alpha_i$'s is captured completely by the
 symmetry factor $|{\rm Aut}(\{\alpha_i\})|=\prod_k m_k!$. {\it In order
 to make full utilization of this fact, it is useful to regard the $\alpha_i$'s as
 elements of a two dimensional vector space spanned by $\alpha$ and
 $\beta$, not necessarily lying on the lattice, and $g_{\rm ref}$ as
 continuous function of these $\alpha_i$'s.} 
 We shall give the expressions
 for $g_{\rm ref}$ for
 generic non-identical, non-parallel
 vectors $\alpha_i$ lying in the first quadrant
 of the two dimensional
 plane spanned by $\alpha$, $\beta$. From this
 we can recover the results for two or more
 identical or parallel $\alpha_i$'s as limits of this general 
 formula.\footnote{For the KS wall crossing
 formula, this prescription was 
proved in \cite{Manschot:2010qz} (last paragraph of \S4.4).
For the MPS wall crossing formul\ae\ this is included in the
prescription for computing $g_{\rm ref}$, and follows from the ability
to replace Bose/Fermi statistics by Boltzmann statistics at the cost
of replacing $\Omega$ by
$\bar\Omega$. To our knowledge this has not been
proved for the Joyce-Song (JS) wall crossing 
formula\cite{MR2357325,Joyce:2008pc,Joyce:2009xv},
but the agreement between JS and other wall crossing
formul\ae\ in explicit examples indicate that this is valid also in
that case.}
 
 The wall crossing formula of Kontsevich and Soibelman 
 (KS)\cite{ks,Kontsevich:2009xt,1006.2706} and 
 Manschot, Pioline and the author (MPS)\cite{Manschot:2010qz} differ in their
 specification of the functions $g_{\rm ref}$. 
 Ref.\cite{Manschot:2010qz} actually proposed two different versions of the wall
 crossing formula. The first one, called the `higgs branch formula', is based on
 Reineke's result on quiver moduli spaces\cite{MR1974891}
 (see also \cite{MR2357325}, \cite{1109.4861} for related results), 
 and the second one, called the
 `coulomb branch
 formula', is based on quantum mechanics of multiple black 
 holes\cite{Denef:2000nb,Denef:2002ru,Bates:2003vx,Denef:2007vg}.
 We shall describe the higgs branch formula for $g_{\rm ref}$ in 
  \S\ref{s3} and the KS formula for $g_{\rm ref}$ in 
\S\ref{s2}.  
 In either case we shall describe the formula as a function
 of generic non-identical, non-parallel  vectors 
 $\alpha_i$ in the first quadrant of the plane spanned by $\alpha$ and 
 $\beta$. 
 The equivalence of the two formul\ae\ 
 was tested in \cite{Manschot:2010qz}
 for low values of $n$
 but was not proven.
 In \S\ref{s4} and \S\ref{s5} we 
prove the equality of these two apprently different formul\ae\
 for $g_{\rm ref}$. Finally in \S\ref{coulomb} we describe the coulomb
 branch formula for $g_{\rm ref}$ and prove its 
 equality with the higgs branch formula.
 
 Since our higgs branch formula is based on Reineke's formula
 on quiver moduli spaces\cite{MR1974891} and since the latter
 has close relationship with both the wall crossing formul\ae\ of
 KS\cite{ks} as well as that of Joyce and 
 Song\cite{MR2357325,Joyce:2008pc,Joyce:2009xv}, the
 equality of the higgs branch formula and the KS formula is
 not unexpected\cite{0804.3214}. Nevertheless our analysis gives
 a direct combinatorial proof of this equivalence. The equivalence
 with the coulomb branch formula is new, -- to our knowledge this
 has not appeared in connection with the wall crossing formula
 before \cite{Manschot:2010qz}.

\noindent{\bf Notations and conventions:} 
We shall end this section by describing some useful notations and conventions
which we shall use.
We define: 
\ben \label{enotations}
\gamma_1<\gamma_2 
&{\rm if}& \langle \gamma_1, \gamma_2\rangle>0 \nn
\gamma_1>\gamma_2 
&{\rm if}& \langle \gamma_1, \gamma_2\rangle<0 \nn
\Theta(\gamma_1, \gamma_2) &=& \cases{1 \quad \hbox{for} \quad \gamma_1>\gamma_2\cr
0  \quad \hbox{for} \quad \gamma_1<\gamma_2} \, . 
\een
Thus for example  if $(\gamma_1, \gamma_2)$ 
follows a
clockwise order then $\gamma_1<\gamma_2$ 
and $\Theta(\gamma_2,\gamma_1)=1$, $\Theta(\gamma_1,\gamma_2)=0$. 
Since this notation will be used extensively in the rest of
the paper, it will be useful to keep in mind the physical picture shown in
Fig.~\ref{f1}. We shall also sometimes describe the situation in Fig.~\ref{f1}
by saying that $\gamma_1$ is to the left of $\gamma_2$ or that
$\gamma_2$ is to the right of $\gamma_1$.
\refb{enotations} satisfies useful identities like:
\ben \label{euseful}
&& \gamma_1< \gamma_2 \Leftrightarrow
\gamma_1< \gamma_1+\gamma_2 \Leftrightarrow
\gamma_1+\gamma_2 < \gamma_2, \qquad
\gamma_1>\gamma_3 \quad \hbox{if} \quad \gamma_1>\gamma_2, 
\quad \gamma_2
> \gamma_3, \nn
&& \Theta(\gamma_1+\gamma_2, \gamma_1)=\Theta(\gamma_2, 
\gamma_1)
= \Theta(\gamma_2, \gamma_1+\gamma_2)\, .
\een
We shall also use the symbol $\Theta$ to
 denote the usual step function of a real variable
 \be \label{ethetareal}
 \Theta(x) = \cases{\hbox{1 for $x\ge 0$}\cr \hbox{0 for $x<0$}}\, .
 \ee
 Which of the two definitions we are using in any given context can be
 understood by examining the argument of $\Theta$.
 
 Since the sum in \refb{efirst3} runs over unordered set of $\alpha_i$'s,
 we can choose a specific order of the $\{\alpha_i\}$ when we give the
 functional form of $g_{\rm ref}$. We shall choose the convention
 in which the $\{\alpha_i\}$'s are ordered as
\be \label{ealphaorder}
\alpha_1<\alpha_2<\alpha_3 \cdots < \alpha_n\, .
\ee
In other words in the two dimensional plane $\alpha_1,\cdots\alpha_n$ form
a clockwise order. We can also express \refb{ealphaorder} as
\be \label{eprop}
\Theta(\alpha_i, \alpha_j) =\Theta(i-j)\, , \qquad \forall \, i,j\,.
\ee
Finally we introduce the shorthand notation
\be \label{eshort}
\alpha_{ij}\equiv  \langle\alpha_i, \alpha_j\rangle\, .
\ee

 In the rest of the paper we shall not explicitly display the variable $y$
 in the argument of
 $g_{\rm ref}$ and other functions, but
 it should be understood that all the quantities depend on $y$.

\sectiono{`Higgs branch' wall crossing formula} \label{s3}

First we shall describe the
 `higgs branch' formula for $g_{\rm ref}$, which will be 
 denoted by $g_{\rm higgs}$.
$g_{\rm higgs}(\alpha_1,\cdots \alpha_n)$ is given by the Poincare polynomial
of a quiver with $n$ nodes, each carrying a U(1) factor, and with
$\alpha_{ij}$ arrows directed from the $i$-th node to the $j$-th node for
$i<j$. The latter in turn is given by the Reineke 
formula\cite{MR1974891}.
The algorithm for calculating $g_{\rm higgs}$ following the original
Reineke formula  leads to many terms whose contributions cancel. 
We shall state the result using a slightly different but equivalent algorithm 
given in \cite{Manschot:2010qz} (\S3.3) where some of these
cancellations are taken into account. Some applications of this
formula can be found in \cite{1106.4238,1110.4847}

Let $\sigma(i)$ for $1\le i\le n$ denote a
permutation of the numbers $1,\cdots n$. 
$g_{\rm higgs}$ is given as a sum over different
permutations $\sigma$. It takes the form:
\ben
\label{higgsfnewest}
 && g_{\rm higgs}(\alpha_1,\dots ,\alpha_n)
 = (-1)^{-1+n} \, \left(y-y^{-1}\right)^{1-n}\, 
 \sum_{\sigma}  {\rm N}^{(n)}_{\rm higgs}(\{\alpha_i\};\sigma)\, (-y)^{
 \sum_{l<k}
 \alpha_{\sigma(l)
 \sigma(k) }}
\nn && \qquad \qquad
 \qquad \quad \, \, \,
=(-1)^{-1+n} (-y)^{
 -\sum_{i<j} \alpha_{ij}} \,
 \left(y-y^{-1}\right)^{1-n} \, \nn && \qquad \qquad
 \qquad \qquad \qquad \times
\sum_{\sigma}  {\rm N}^{(n)}_{\rm higgs}(\{\alpha_i\};\sigma)\, (-y)^{
 2\sum_{l<k\atop\sigma(l)<\sigma(k)}
 \alpha_{\sigma(l)
 \sigma(k) }}\, , 
 \nn
 && {\rm N}^{(n)}_{\rm higgs}(\{\alpha_i\};\sigma)  \nn &=& 
 (-1)^{s(\sigma)-1}
 \prod_{k=2\atop \sigma(k)<\sigma(k-1)}^n \Theta\left( \alpha_1+
 \cdots \alpha_n,\sum_{i=k}^n \alpha_{\sigma(i)}\right)
 \prod_{k=2\atop \sigma(k) > \sigma(k-1)}^n \Theta\left(
 \sum_{i=k}^n \alpha_{\sigma(i)},\alpha_1+\cdots \alpha_n\right)
 \nn
 &=&(-1)^{s(\sigma)-1}
 \prod_{k=1\atop \sigma(k+1)<\sigma(k)}^{n-1} \Theta\left(\sum_{i=1}^{k} 
 \alpha_{\sigma(i)}, \alpha_1+
 \cdots \alpha_n\right)
 \prod_{k=1\atop \sigma(k+1) > \sigma(k)}^{n-1} \Theta\left(
 \alpha_1+\cdots \alpha_n,
 \sum_{i=1}^{k} \alpha_{\sigma(i)}\right)
 \, , \nn
 &s(\sigma)&= 1+\sum_{k=1}^{n-1} 
 \Theta(\alpha_{\sigma(k)}, \alpha_{\sigma(k+1)})
 =  1+\sum_{k=1}^{n-1} \Theta(\sigma(k) -\sigma(k+1))\, .
\een
The  $\Theta$ 
in the second expression for $s(\sigma)$ is the
ordinary step function.
It has been shown in appendix \ref{sa} that this is equivalent to the
formula derived in \S3.3 of \cite{Manschot:2010qz} which in turn was shown in
\cite{Manschot:2010qz} to be equivalent to the
Reineke formula\cite{MR1974891}.  The equality of the
two expressions for
${\rm N}^{(n)}_{\rm higgs}(\{\alpha_i\};\sigma)$ given in
\refb{higgsfnewest} 
follows from a simple shift $k\to k+1$ and the
identities given in \refb{euseful}.  Although for physical charges the
$\alpha_{ij}$'s are integers and hence \refb{higgsfnewest} is uniquely
defined everywhere in the complex $y$-plane, we shall at the intermediate
steps work with analytically continued $\alpha_{ij}$'s away from integer
values. In this case we shall use \refb{higgsfnewest} to define
$g_{\rm higgs}$ along the negative $y$-axis in the range $-1<y<0$
and then analytically
continue the result to the rest of the complex plane.\footnote{A physical
interpretation of the  exponent $ \sum_{l<k}
 \alpha_{\sigma(l)
 \sigma(k) }$ is as follows. Let us represent the $\alpha_i$'s as vectors in the
 two dimensional plane such that $\alpha_{ij}$ is the area of the
 parallelogram with sides $\alpha_i$ and $\alpha_j$. Then
 $\sum_{l<k}
 \alpha_{\sigma(l)
 \sigma(k) }$ is the area of the oriented polygon with sides 
 $\alpha_{\sigma(1)}$, $\alpha_{\sigma(2)}$, $\cdots$ $\alpha_{\sigma(n)}$
 and $-\alpha_{\sigma(1)}-\cdots -\alpha_{\sigma(n)}$.
 In general the polygon can be self-intersecting in which case the area has
 to be taken as the sum of the areas of each component polygon
 weighted with $\pm 1$ depending on the orientation of the boundary of
 that particular
 component. I wish to
 thank the referee for suggesting this interpretation.
}

Since \refb{higgsfnewest} will play a central role in our analysis, it will be
useful to keep in mind a physical picture of this equation. What this equation tells
us is that for a given permutation to contribute to $g_{\rm higgs}$ it must
satisfy the conditions:
\ben \label{eexplicit}
\sum_{i=1}^{k} 
 \alpha_{\sigma(i)} &<& \alpha_1+
 \cdots \alpha_n \quad \hbox{for $\sigma(k)<\sigma(k+1)$} \nn
 \sum_{i=1}^{k} 
 \alpha_{\sigma(i)} &>& \alpha_1+
 \cdots \alpha_n \quad \hbox{for $\sigma(k)>\sigma(k+1)$} \, .
 \een
 Furthermore, when the above condition is satisfied, its contribution to
 ${\rm N}^{(n)}_{\rm higgs}$ is 1 or $-1$ depending on whether the number of neighboring
 pairs for which $\sigma(i)>\sigma(i+1)$ is even or odd.

For $n=2$ the permutations are $(12)$ and $(21)$. 
Using \refb{higgsfnewest} we get $s(12)=1$, $s(21)=2$,
${\rm N}^{(2)}_{\rm higgs}(12)=
(-1)^{s(12)-1}\Theta(\alpha_1+\alpha_2,
\alpha_1)=\Theta(\alpha_2,\alpha_1)=1$ and 
${\rm N}^{(2)}_{\rm higgs}(21)=(-1)^{s(21)-1}
\Theta(\alpha_2,\alpha_1+\alpha_2)
=-\Theta(\alpha_2,\alpha_1)
=-1$.
Thus  we get
\be \label{egmps2}
g_{\rm higgs}(\alpha_1,\alpha_2) = -(-y)^{-\alpha_{12}} 
(y-y^{-1})^{-1} ((-y)^{2\alpha_{12}}-1)\, .
\ee

We shall end this section 
by summarizing some useful properties of $g_{\rm higgs}$:
\begin{enumerate}
\item $g_{\rm higgs}$ contains a sum of exponents of the form
$(-y)^{\sum_{i<j} {s_{ij}}\alpha_{ij}}$ where $s_{ij}=1$ or
$-1$. Since there are $n(n-1)/2$ pairs
of $\alpha_{ij}$'s, there are $2^{n(n-1)/2}$ possible choices
of the $\{s_{ij}\}$'s.
However of these only those terms which have the form:
\be \label{epr1}
(-y)^{\sum_{i<j\atop \sigma(i)<\sigma(j)} \alpha_{\sigma(i)\sigma(j)}
- \sum_{i>j\atop \sigma(i)<\sigma(j)} \alpha_{\sigma(i)\sigma(j)}}
= (-y)^{\sum_{i<j} \alpha_{\sigma(i)\sigma(j)}}
\ee
for some permutation $\sigma$
appear in the sum. This already restricts the sum to $n!$ terms
corresponding to $n!$ possible choices of $\sigma$. The
constraints \refb{eexplicit} further reduce
the number of terms.

\item Let $A_a$ ($1\le a\le 2^n-1$) denote the collection of all 
non-empty subsets of $\{1,2,\cdots n\}$ and let 
\be \label{edefgamma}
\gamma^{(a)}
\equiv \sum_{i\in A_a} \alpha_i\, .
\ee 
Then ${\rm N}^{(n)}_{\rm higgs}(\{\alpha_i\}; \sigma)$
depends only on the relative orientation of the vectors $\gamma^{(a)}$
relative to $(\alpha_1+\cdots \alpha_n)$, but not on the relative
orientations of $\gamma^{(a)}$ and $\gamma^{(b)}$. This is apparent from
the fact that the argument of the $\Theta$'s appearing in \refb{higgsfnewest}
involve only the pairs $(\gamma^{(a)}, \alpha_1+\cdots \alpha_n)$
but not $(\gamma^{(a)}, \gamma^{(b)})$.

\item We can improve upon the above result if we focus on the term
corresponding to a given permutation $\sigma$. The corresponding
${\rm N}^{(n)}_{\rm higgs}(\{\alpha_i\}; \sigma)$ depends only on the relative orientation
of $\sum_{i=1}^k \alpha_{\sigma(i)}$ and $\alpha_1+\cdots \alpha_n$
for $1\le k\le n-1$, and of course the relative orientation of the
pairs $(\alpha_i, \alpha_j)$.
All the other $\gamma^{(a)}$'s are irrelevant. Thus while computing
${\rm N}^{(n)}_{\rm higgs}(\{\alpha_i\}; \sigma)$ for a particular
$\sigma$ we can freely deform the $\alpha_i$'s
as long as we do not change the relative orientation between
$\sum_{i=1}^k \alpha_{\sigma(i)}$ and $\alpha_1+\cdots \alpha_n$
for any $k$, and  also preserve the relative orientation between
the $\alpha_i$'s.

\item 
If a permutation $\sigma$ appears in the sum in \refb{higgsfnewest}, 
then the
permutation $\sigma'$ where the order of all the elements is reversed, also
appears in the sum.

\noindent Proof: 
We have
\be \label{esigmapdef}
\sigma'(i) = \sigma(n+1-i)\, .
\ee
Eq.\refb{higgsfnewest} now gives
\ben \label{eproofrev}
{\rm N}^{(n)}_{\rm higgs}(\{\alpha_i\};\sigma')  &=&(-1)^{s(\sigma')-1}
 \prod_{k=1\atop \sigma'(k+1)<\sigma'(k)}^{n-1} \Theta\left(\sum_{i=1}^{k} 
 \alpha_{\sigma'(i)}, \alpha_1+
 \cdots \alpha_n\right) \nn && \qquad \qquad 
 \prod_{k=1\atop \sigma'(k+1) > \sigma'(k)}^{n-1} \Theta\left(
 \alpha_1+\cdots \alpha_n,
 \sum_{i=1}^{k} \alpha_{\sigma'(i)}\right)
 \, , \nn
 &=& (-1)^{s(\sigma')-1}
 \prod_{k=1\atop \sigma(n-k)<\sigma(n-k+1)}^{n-1} \Theta\left(\sum_{i=1}^{k} 
 \alpha_{\sigma(n-i+1)}, \alpha_1+
 \cdots \alpha_n\right) \nn && \qquad \qquad 
 \prod_{k=1\atop \sigma(n-k) > \sigma(n-k+1)}^{n-1} \Theta\left(
 \alpha_1+\cdots \alpha_n,
 \sum_{i=1}^{k} \alpha_{\sigma(n-i+1)}\right)
 \, , \nn
 &=& (-1)^{s(\sigma')-1}
 \prod_{\ell=1\atop \sigma(\ell)<\sigma(\ell+1)}^{n-1} \Theta\left(\sum_{j=
 \ell+1}^{n} 
 \alpha_{\sigma(j)}, \alpha_1+
 \cdots \alpha_n\right) \nn && \qquad \qquad 
 \prod_{\ell = 1\atop \sigma(\ell)>\sigma(\ell+1)}^{n-1} \Theta\left(
 \alpha_1+\cdots \alpha_n,
 \sum_{j=\ell+1}^{n} \alpha_{\sigma(j)}\right)
 \, ,
 \een
 with
 \ben \label{essigrev}
 s(\sigma')= 1+\sum_{k=1}^{n-1} \Theta(\sigma'(k) -\sigma'(k+1))=
  1+\sum_{k=1}^{n-1} \Theta(\sigma(n-k+1) -\sigma(n-k)) \nn
  = 1+\sum_{j=1}^{n-1} \Theta(\sigma(j+1) -\sigma(j)) 
 = 1+ (n-1) - \sum_{j=1}^{n-1} \Theta(\sigma(j) -\sigma(j+1))\, .
 \een
 Comparing \refb{eproofrev}, \refb{essigrev} with \refb{higgsfnewest}
 we get
 \be \label{erevfin}
 {\rm N}^{(n)}_{\rm higgs}(\{\alpha_i\};\sigma') = (-1)^{n-1}  {\rm N}^{(n)}_{\rm higgs}(\{\alpha_i\};\sigma)\, ,
 \ee
 showing that ${\rm N}^{(n)}_{\rm higgs}(\{\alpha_i\};\sigma')$ is non-zero iff
 ${\rm N}^{(n)}_{\rm higgs}(\{\alpha_i\};\sigma)$ is non-zero.
 Since reversing the permutation reverses the sign of
$\sum_{i<j} \alpha_{\sigma(i)\sigma(j)}$, the result given above shows that
$g_{\rm higgs}$ is invariant under $y\to y^{-1}$.
This is of course expected from the fact that $g_{\rm higgs}$ is the Poincare
polynomial of the moduli space of abelian quivers.

\end{enumerate}

Finally note that if we are interested in
the ordinary (numerical)
index instead of the refined index, the relevant $g$ is obtained by
taking the $y\to 1$ limit of 
\refb{higgsfnewest}\cite{1110.4847}. 
This limit is apparently singular, but given that
$g_{\rm higgs}(\alpha_1,\dots ,\alpha_n)$ is the Poincare polynomial
of abelian quivers and hence has a finite $y\to 1$ limit, the singularities
must cancel after we sum over all permutations $\sigma$. Thus 
if we define $y=e^\nu$ and expand 
$\sum_{\sigma}  {\rm N}^{(n)}_{\rm higgs}(\{\alpha_i\};\sigma)\, (-y)^{
 \sum_{l<k}
 \alpha_{\sigma(l)
 \sigma(k) }}$
 in a power series in $\nu$, all powers of $\nu$ up to $\nu^{n-2}$
 must cancel.  As a result 
we can
extract the $y\to 1$ limit of $g_{\rm higgs}$ by picking the
order $\nu^{n-1}$ term from the expansion of each
$(-y)^{
 \sum_{l<k}
 \alpha_{\sigma(l)
 \sigma(k) }}$ term, and then taking the $\nu\to 0$ limit 
 of the resulting expression.
This gives, 
\ben \label{eordinary}
g_{\rm numerical}(\alpha_1,\dots ,\alpha_n)
\qquad \qquad \qquad \qquad \qquad \qquad \qquad \qquad 
\qquad \qquad \qquad \qquad \qquad \nn
 = (-1)^{-1+n }\, 2^{1-n}\, {1\over (n-1)!}\,  \times
\sum_{\sigma}  {\rm N}^{(n)}_{\rm higgs}(\{\alpha_i\};\sigma)
(-1)^{\sum_{l<k}
 \alpha_{\sigma(l)
 \sigma(k) }}\left( \sum_{l<k}
 \alpha_{\sigma(l)
 \sigma(k) }
\right)^{n-1}\nn
 \to (-1)^{-1+n }\, 2^{1-n}\, {1\over (n-1)!}\,  (-1)^{\sum_{l<k}
 \alpha_{lk }}\times
\sum_{\sigma}  {\rm N}^{(n)}_{\rm higgs}(\{\alpha_i\};\sigma)
\left( \sum_{l<k}
 \alpha_{\sigma(l)
 \sigma(k) }
\right)^{n-1} \, , \quad
\een
where the second expression is valid in the limit when the $\alpha_i$'s,
instead of being general two dimensional vectors, approach
lattice vectors so that $\alpha_{ij}$'s approach integers. 
For generic charges
\refb{eordinary} appears to be closely related to, but  not
quite the same as the JS wall crossing 
formula\cite{MR2357325,Joyce:2008pc,Joyce:2009xv}.
(A short review of the JS formula and its implementation can be found
in \cite{Manschot:2010qz}, section 5.)
In particular in the JS formula the summand $\left( \sum_{l<k}
 \alpha_{\sigma(l)
 \sigma(k) }
\right)^{n-1} $ is replaced by a slightly different term obtained by summing
over trees.
However for non-generic charges
\i.e.\ when some $\alpha_i$'s -- and/or their linear
combinations with positive integer coefficients -- are equal or parallel
to each other, the JS prescription involves sum over many more
terms, while the MPS prescription simply requires us to take the limit
of the formula for generic charges and supply the Boltzmann factor
$1/|{\rm Aut}(\{\alpha_i\})|=
1/\prod_k m_k!$ as described in \refb{efirst3}.
It will be interesting to find
a direct combinatoric proof of the equivalence of \refb{eordinary}
with the JS wall crossing formula.

\sectiono{KS wall crossing formula} \label{s2}

We shall now describe a version
of the KS wall crossing formula given in
\cite{Manschot:2010qz}. To describe the KS wall crossing
 formula we introduce an algebra with elements of the form $e_\gamma$
 with $\gamma\in \Lambda$, satisfying the commutation relations: 
\be \label{eliealg}
[ e_{\gamma} ,  e_{\gamma'}] = \kappa(\gamma, \gamma') \,
 e_{\gamma+\gamma'}, \qquad \kappa(\gamma, \gamma') =
{ (-y)^{\langle\gamma, \gamma' \rangle}
- (-y)^{- \langle\gamma, \gamma' \rangle}\over y- y^{-1}}\, 
\, \ .
\ee 
Let $\alpha_1, \cdots \alpha_n$ be a set of vectors arranged so that
$\alpha_1<\alpha_2<\cdots < \alpha_n$, \i.e.\
in the two dimensional representation $\alpha_1,\cdots \alpha_n$ are
arranged in a clockwise fashion.
As before, we denote by $\{A_a\}$ the collection of all possible non-empty
subsets of the integers $1,\cdots n$, and define
$\gamma^{(a)}=\sum_{i\in A^{(a)}}\alpha_i$. We shall order the $A_a$'s
so that $\gamma^{(a)}$'s form a clockwise order 
as $a$ increases: $\gamma^{(a)}<\gamma^{(b)}$ for $a<b$.
Now we begin with the product
$e_{\alpha_n}\cdots e_{\alpha_1}$ and then try to reverse the order using \refb{eliealg},  
bringing this into a linear combination of terms of the form
$e_{\gamma^{(a_1)}} e_{\gamma^{(a_2)}}\cdots 
e_{\gamma^{(a_k)}}$ with $a_1<a_2<\cdots < a_k$, 
$\gamma^{(a_1)}+\cdots \gamma^{(a_k)} = \alpha_1+\cdots \alpha_n$:
\be \label{eksreverse}
e_{\alpha_n}\cdots e_{\alpha_1}= \sum_{k=1}^n \sum_{\{a_1,\cdots a_k\}\atop
\gamma^{(a_1)}+\cdots \gamma^{(a_k)}=\alpha_1+\cdots +\alpha_n;
a_1<a_2\cdots <a_k}
h(\alpha_1,\cdots \alpha_n; \gamma^{(a_1)}, \cdots
\gamma^{(a_k)}) \, e_{\gamma^{(a_1)}} \cdots e_{\gamma^{(a_k)}} \, ,
\ee
for some functions $h$.
The $g_{\rm ref}$ for KS wall crossing formula, denoted by
$g_{\rm KS}$, is given by the coefficient of $e_{\alpha_1+\cdots +\alpha_n}$
in this expression:
\be \label{eksfin}
g_{\rm KS}(\alpha_1,\cdots \alpha_n) =
h(\alpha_1, \cdots \alpha_n; \alpha_1+\cdots \alpha_n)\, .
\ee
For example for $n=2$ we write
\be \label{en2ks}
e_{\alpha_2}e_{\alpha_1} = e_{\alpha_1} e_{\alpha_2} + 
\kappa(\alpha_2, \alpha_1) \, e_{\alpha_1+\alpha_2}\, .
\ee
Thus we have 
\be \label{eas1}
g_{\rm KS}(\alpha_1, \alpha_2) = \kappa(\alpha_2, \alpha_1) 
\, .
\ee
This agrees with the corresponding formula \refb{egmps2} for 
$g_{\rm higgs}(\alpha_1, \alpha_2)$.
The equivalence of $g_{\rm KS}(\alpha_1, \cdots \alpha_n)$ 
and $g_{\rm higgs}(\alpha_1,
\cdots \alpha_n)$ has been tested explicitly up to $n\le 5$\cite{Manschot:2010qz}.

We shall now examine if $g_{\rm KS}$ also satisfies the four properties
of $g_{\rm higgs}$ listed at the end of \S\ref{s3}.
\begin{enumerate}
\item 
We shall first show that like in the
expression for $g_{\rm higgs}$ given in \refb{higgsfnewest},
each term in
$g_{\rm KS}$ can also be associated with a permutation, \i.e.\
$g_{\rm KS}$ can be expressed as
\ben\label{egksperm}
 g_{\rm KS}(\alpha_1,\dots ,\alpha_n)\qquad \qquad \qquad \qquad \qquad 
 \qquad \qquad \qquad \qquad  \qquad \qquad \qquad \qquad \nn
= (-1)^{-1+n} (-y)^{
 -\sum_{i<j} \alpha_{ij}} \,
 \left(y-y^{-1}\right)^{1-n} 
\sum_{\sigma}  {\rm N}^{(n)}_{\rm KS}(\{\alpha_i\};\sigma) (-y)^{
 2\sum_{l<k\atop\sigma(l)<\sigma(k)}
 \alpha_{\sigma(l)
 \sigma(k) }} \nn
 = (-1)^{-1+n}  \,
 \left(y-y^{-1}\right)^{1-n} 
\sum_{\sigma}  {\rm N}^{(n)}_{\rm KS}(\{\alpha_i\};\sigma) (-y)^{
 \sum_{l<k}
 \alpha_{\sigma(l)
 \sigma(k) }}\qquad \qquad \qquad \qquad 
 \een
 for some integers ${\rm N}^{(n)}_{\rm KS}(\{\alpha_i\};\sigma)$.
Suppose we begin with a pair of generators $e_{\alpha_i}$,
$e_{\alpha_j}$ and pick up their commutator. Then the coefficient
of this term, besides the $(y-y^{-1})^{-1}$ term, is
proportional to $(-y)^{\alpha_{ij}} - (-y)^{-\alpha_{ij}}$. The
first term has the interpretation of a permutation in which
$i$ is to the left of $j$ and the second term has the interpretation
of being associated with a permutation in which $j$ appears to the
left of $i$. If we now pick the commutator of $e_{\alpha_i+\alpha_j}$
with a third generator $e_{\alpha_k}$, then we get a factor proportional to
$(-y)^{\alpha_{ik}+\alpha_{jk}} - (-y)^{-\alpha_{ik}-\alpha_{jk}}$.
The first term has the interpretation of a permutation in which
the $i$ and $j$ are to the left of $k$ and the second term has the
interpretation of a permutation in which $k$ is to the left of
$i$ and $j$. 
Thus this can be combined with the earlier ordering $(ij)$ or $(ji)$
without any conflict.\footnote{In contrast if we had found a term like
$\alpha_{ik}-\alpha_{jk}-\alpha_{ij}$ in the exponent it would have
the interpretation that $k$ is to the right of $i$ and left of $j$, and $j$
is to the left of $i$. Clearly 
there is no arrangement of $i,j,k$ satisfying these requirements.}
This argument can be extended to more general situations.
Individual steps in arriving at \refb{eksreverse} consists of manipulating a
product $e_{\gamma^{(a)}}e_{\gamma^{(b)}}$ by reversing their
order
where
$\gamma^{(a)}$ and $\gamma^{(b)}$ are defined as in
\refb{edefgamma} with non-overlapping sets $A_a$ and $A_b$.
Now while reversing the order in the product 
$e_{\gamma^{(a)}}e_{\gamma^{(b)}}$
the 
commutator gives $e_{\gamma^{(a)}+\gamma^{(b)}}$
multiplied by a factor proportional to
$(-y)^{\langle \gamma^{(a)}, \gamma^{(b)}\rangle} -
(-y)^{\langle \gamma^{(b)}, \gamma^{(a)}\rangle}$. The first term can be
regarded as coming from a configuration where all the elements
of $A_a$ are to the left of all the elements of $A_b$ and the second
term can be regarded as coming from a configuration where all the
elements of $A_a$ are to the right of all the elements of $A_b$.
Suppose further that earlier, while arriving at 
$e_{\gamma^{(a)}}$ by combining $e_{\alpha_i}$ for $i\in A_a$
we have gotten a sum of terms
each of which can be associated with the permutation of the elements
inside $A_a$ and a similar relation holds for $e_{\gamma^{(b)}}$.
When we multiply these by the $(-y)^{\langle \gamma^{(a)}, \gamma^{(b)}\rangle} -
(-y)^{\langle \gamma^{(b)}, \gamma^{(a)}\rangle}$
factor, individual terms
in the product will correspond
to specific permutation of the elements inside $A_a$ and specific
permutation of the elements inside $A_b$, and on top of that 
all the elements
of $A_a$ could be to the left of all the elements of $A_b$ or all the
elements of $A_b$  could be to the left of all the elements of $A_a$.
Thus each term multiplying $e_{\gamma^{(a)}+\gamma^{(b)}}$
can be regarded as coming from some permutation of the elements
of $A_a\cup A_b$.
This now shows by induction that
at every stage of the manipulation that leads to the KS formula
for wall crossing, we produce a set of terms each of which can be associated
with a permutation of the $\alpha_i$'s involved. As a result 
the final expression
for $g_{\rm KS}$ must also
contain only those powers of $y$ which have the interpretation
of being associated with a permutation as in \refb{egksperm}.
We shall see this more explicitly in \refb{emps2}, \refb{emps3},
\refb{emps4}.

\item Like $g_{\rm higgs}$, $g_{\rm KS}$ is also a piecewise analytic function
of the $\alpha_i$'s. The form of the function depends on the relative orientation
between $\gamma^{(a)}$ and 
$\alpha_1+\cdots \alpha_n$ and is independent of the relative orientation
between the $\gamma^{(a)}$'s. This was proved in
\cite{Manschot:2010qz}, but for completeness we shall repeat the proof.
Let us suppose that by manipulating the product $e_{\alpha_n}\cdots
e_{\alpha_1}$ we have brought it into the form \refb{eksreverse}.
Since $\kappa(\gamma, \gamma')$ is an analytic function of $\gamma$,
$\gamma'$, it follows from \refb{eliealg} that the coefficient of
$e_{\alpha_1+\cdots \alpha_n}$  in this expression, which is given by 
sum of products of
$\kappa(\gamma,\gamma')$ with $\gamma=\sum_{i\in A}\alpha_i$,
$\gamma'=\sum_{i\in B} \alpha_i$ for some subsets $A$ and $B$
of $\{1,2,\cdots n\}$, is an analytic function of
the $\alpha_i$'s inside a chamber in which the relative order of the
$\gamma^{(a)}$'s is fixed.
Now suppose that we deform some of the $\alpha_i$'s to make a
pair of 
$\gamma^{(a)}$'s 
switch their relative orientation but none of the
$\gamma^{(a)}$'s cross the ray corresponding to $\alpha_1+\cdots
\alpha_n$. 
In particular let us suppose that the relevant pairs are 
$\gamma^{(b)}$ and $\gamma^{(c)}$, and that the deformation takes
us from $\gamma^{(b)}<\gamma^{(c)}$ to $\gamma^{(b)}>\gamma^{(c)}$.
In this case
$\gamma^{(b)}$ and $\gamma^{(c)}$ 
are either both on the left or both on the
right of $\alpha_1+\cdots \alpha_n$.
Let us for definiteness assume that they are both to the left. Now to
bring the products of $e_{\gamma^{(a)}}$'s to the standard order we need to
express the product $e_{\gamma^{(b)}} e_{\gamma^{(c)}}$ inside any
term as $e_{\gamma^{(c)}} e_{\gamma^{(b)}}+\kappa(\gamma^{(b)}, \gamma^{(c)})
e_{\gamma^{(b)}+\gamma^{(c)}}$. However since $\gamma^{(b)}$ and $\gamma^{(c)}$
are both on the left of $\alpha_1+\cdots \alpha_n$, the extra term
proportional to $e_{\gamma^{(b)}+\gamma^{(c)}}$ can never give a factor of
$e_{\alpha_1+\cdots \alpha_n}$. Thus under such deformations 
$g_{\rm KS}(\alpha_1, \cdots \alpha_n)$, which is the coefficient of
$e_{\alpha_1+\cdots \alpha_n}$ at the end of this manipulation, remains
unchanged. This shows that $g_{\rm KS}(\alpha_1, \cdots \alpha_n)$ is a
piecewise analytic function of the $\alpha_i$'s, with the form of the function
determined by the relative orientation between the $\gamma^{(a)}$'s
and $\alpha_1+\cdots \alpha_n$.
\item As in the case of $g_{\rm higgs}$, one can improve the result if we
focus on a term corresponding to a given permutation $\sigma$.
We shall show that
in this case the coefficient depends only on the relative orientation of
$\sum_{i=1}^k \alpha_{\sigma(i)}$ and $\alpha_1+\cdots \alpha_n$
for different values of $k$, and not
on the relative orientation between $\gamma^{(a)}$ and
$\alpha_1+\cdots \alpha_n$ for other $\gamma^{(a)}$'s. To see this let us
again suppose that by manipulating the product $e_{\alpha_n}\cdots
e_{\alpha_1}$ we have brought it into the form \refb{eksreverse}. Now
suppose that we deform the $\alpha_i$'s so that a specific $\gamma^{(a)}$
crosses $\alpha_1+\cdots \alpha_n$ from  left to right. At the same time the 
vector $\gamma^{(b)}
\equiv \alpha_1+\cdots \alpha_n -\gamma^{(a)}$ crosses 
$\alpha_1+\cdots \alpha_n$ from right to left. Before deformation the
generators $e_{\gamma^{(a)}}$ and
$e_{\gamma^{(b)}}$
would have been arranged in the order
$e_{\gamma^{(a)}} e_{\gamma^{(b)}}$, but after the deformation we
need to reverse their order picking up a term proportional to
$\{(-y)^{\langle \gamma^{(a)}, \gamma^{(b)}\rangle} -
(-y)^{\langle \gamma^{(b)}, \gamma^{(a)}\rangle}\} 
e_{\alpha_1+\cdots \alpha_n}$. The first term has
the interpretation of all the elements in the set $A_a$ being to the left of
all the elements in the set $A_b$ (which is the complement of the set
$A_a$) and the second term
has
the interpretation of all the elements in the set $A_b$ being to the left of
all the elements in the set $A_a$.
Thus such a term can change the coefficient of a term associated with the
permutation $\sigma$ only if in $\sigma$ 
all the elements of $A_a$ are
to the left (or right) of all the elements of the compliment of
$A_a$.  In other words $A_a$ must contain a set of $k$ elements
to the left (or a set of $(n-k)$ elements to the right) for some integer
$k$. This shows that 
the coefficient of a term associated with the permutation
$\sigma$ in $g_{\rm KS}$ can only depend on the relative orientation 
between $\alpha_{\sigma(1)}+\cdots \alpha_{\sigma(k)}$ 
and
$\alpha_1+\cdots \alpha_n$ for different
integers $k$ but not on the relative orientation between
$\gamma^{(a)}$ and $\alpha_1+\cdots \alpha_n$ for other sets
$A_a$.

\item Finally we turn to the fourth property of $g_{\rm higgs}$ which states that
if a permutation $\sigma$ appears in $g_{\rm higgs}$ then its reverse
permutation will also appear. As discussed at the end of \S\ref{s3},
this is equivalent to proving the symmetry of $g_{\rm higgs}$ under
$y\to y^{-1}$. This property is automatic in $g_{\rm KS}$ since the $y$
dependence arises from the $\kappa(\gamma_1, \gamma_2)$ 
factors which are
manifestly invariant under $y\to y^{-1}$.

\end{enumerate}

In the next two sections we shall prove the equality of
$g_{\rm KS}(\alpha_1, \cdots \alpha_n)$ and 
$g_{\rm higgs}(\alpha_1, \cdots \alpha_n)$ for generic charge
vectors $\{\alpha_i\}$ for which all
vectors of the form $\sum_{i\in A} \alpha_i$ for different 
subsets $A$ of
$\{1, \cdots n\}$ are strictly ordered. The special cases
where some of these vectors are parallel or identical to each
other are then obtained as limts of this generic case. One question that
arises naturally is whether the limit is well defined, \i.e.\ whether it could
depend on which side we approach the limit from. This is somewhat
obscure for the expression
\refb{higgsfnewest} for $g_{\rm higgs}$ since it 
involves the step functions which jump discontinuously
as the relative ordering between the vectors in the argument switch.
However for $g_{\rm KS}$ defined through \refb{eksreverse}, 
\refb{eksfin} it is clear that the limit is well defined, since the effect of
switching the order between two vectors $\gamma$ and $\gamma'$
vanishes as $\gamma$ and $\gamma'$ become parallel to each other:
$e_\gamma e_{\gamma'} \to e_{\gamma'} e_\gamma$ when
$\gamma$ and $\gamma'$ are parallel. 
Thus for example if we approach a configuration where 
two vectors $\alpha_i$ and $\alpha_j$ become
parallel to each other, then the left hand side of \refb{eksreverse}
is independent of whether they
approach this configuration from the $\alpha_i>\alpha_j$ side
or $\alpha_i<\alpha_j$ side. Similarly if a pair of
$\gamma^{(a)}$'s become parallel to each other, then $h(\alpha_1,\cdots \alpha_n;
\alpha_1+\cdots \alpha_n)$ appearing on the right hand side of
\refb{eksreverse} is independent of how this limit is approached.
Eq.\refb{eksfin} then shows that for $g_{\rm KS}$ the limit to degenerate
configuration of vectors is well defined.
The equality of $g_{\rm KS}$
and $g_{\rm higgs}$ for generic vectors, which will be proved in the next two
sections, 
then implies that
even for $g_{\rm higgs}$ the limit to degenerate configurations of vectors
is well defined, \i.e.\ it does not depend on which side we take the
limit from.

We must reemphasize however 
that for degenerate configurations of vectors
$g_{\rm higgs}$ must be defined as a limit of \refb{higgsfnewest} for
non-degenerate configurations. For example if we were to write a
computer program for computing $g_{\rm higgs}$, the algorithm must
involve adding to the $\alpha_i$'s some randomly generated two dimensional
vectors of small magnitude -- which makes the configuration
non-degenerate -- while computing $N^{(n)}_{\rm higgs}(\{\alpha_i\};
\sigma)$, but while computing the exponent $\sum_{l<k}\alpha_{\sigma(l)\sigma(k)}$
of $(-y)$ we can continue to use the original vectors. For a fixed permutation $\sigma$
the quantities $N^{(n)}_{\rm higgs}(\{\alpha_i\};
\sigma)$ may depend on the choice of the random vectors which we add to
the $\alpha_i$'s, but the argument of the previous paragraph shows that
the final result for $g_{\rm higgs}(\alpha_1,\cdots \alpha_n)$ will be
independent of this choice.

\sectiono{Recursion relations for the KS wall crossing formula} \label{s4} 

We shall now derive a set of recursion relations for $g_{\rm KS}(\alpha_1,
\cdots \alpha_n)$.
Since $\gamma^{(a)}=
\sum_{i\in A_a} \alpha_i$, it is clear that the $e_{\gamma^{(a)}}$
factor in \refb{eksreverse} arises as a result of manipulating the
product of $e_{\alpha_i}$'s for $i\in A_a$, to bring it from the anti-clockwise
ordering to the clockwise ordering. Furthermore the result of this manipulation
is not affected by the $\alpha_i$'s outside the set $A_a$, and hence
gives a factor of $g_{\rm KS}(\{\alpha_i, i\in A_a\})$. Thus 
we get
\be \label{eget1}
h(\alpha_1,\cdots \alpha_n; \gamma^{(a_1)}, \cdots
\gamma^{(a_k)}) = \prod_{\ell=1}^k g_{\rm KS}(\{\alpha_i, \, i\in A_{a_\ell}\})\, .
\ee
Using this we may rewrite \refb{eksreverse} as
\be \label{eksreversenew}
e_{\alpha_n}\cdots e_{\alpha_1}= \sum_{k=1}^n
\sum_{\{a_1,\cdots a_k\}\atop
\gamma^{(a_1)}+\cdots \gamma^{(a_k)}=\alpha_1+\cdots +\alpha_n;
a_1<a_2\cdots <a_k}
\left(\prod_{\ell=1}^k g_{\rm KS}(\{\alpha_i, \, i\in A_{a_\ell}\}) \right)
\, e_{\gamma^{(a_1)}} \cdots e_{\gamma^{(a_k)}} \, .
\ee

We shall now use \refb{eksreversenew} to derive a recursive procedure
for determining $g_{\rm KS}$. Suppose we know the result for $g_{\rm KS}(\alpha_1,
\cdots \alpha_n)$.
Then to find $g_{\rm KS}(\alpha_1,\cdots \alpha_n, \alpha_{n+1})$ with
$\alpha_1<\alpha_2< \cdots <\alpha_n<\alpha_{n+1}$,
we multiply eq.\refb{eksreversenew} from the left by $e_{\alpha_{n+1}}$,
and then try to rearrange the right hand side by moving $e_{\alpha_{n+1}}$ to
the extreme right, so that in each product 
the $e_\gamma$'s have their $\gamma$'s
in clockwise order as we move from left to right. 
For example in the first step
we write
\be \label{eks1}
e_{\alpha_{n+1}} e_{\gamma^{(a_1)}} e_{\gamma^{(a_2)}} 
\cdots e_{\gamma^{(a_k)}}= 
e_{\gamma^{(a_1)}} e_{\alpha_{n+1}} 
e_{\gamma^{(a_2)}} \cdots e_{\gamma^{(a_k)}}+ \kappa(\alpha_{n+1},
\gamma^{(a_1)}) e_{\gamma^{(a_1)}+ \alpha_{n+1}}
e_{\gamma^{(a_2)}} \cdots e_{\gamma^{(a_k)}}\, .
\ee  
In the next step we manipulate the first term as
\ben \label{eks2}
&&e_{\gamma^{(a_1)}} e_{\alpha_{n+1}} e_{\gamma^{(a_2)}}
e_{\gamma^{(a_3)}} \cdots e_{\gamma^{(a_k)}}\nn
&=& e_{\gamma^{(a_1)}}  e_{\gamma^{(a_2)}}  e_{\alpha_{n+1}} 
e_{\gamma^{(a_3)}} \cdots e_{\gamma^{(a_k)}}
+ \kappa(\alpha_{n+1},
\gamma^{(a_2)}) e_{\gamma^{(a_1)}}  
e_{\gamma^{(a_2)}+ \alpha_{n+1}}e_{\gamma^{(a_3)}} \cdots e_{\gamma^{(a_k)}}\, .
\een
For the second term of \refb{eks1}  we have to consider
two possibilities. If
$\gamma^{(a_1)}+ \alpha_{n+1} < \gamma^{(a_2)}$ we already have all
the terms in the product
in the correct order and we can stop manipulating this term.
On the other hand if 
$\gamma^{(a_1)}+ \alpha_{n+1} > \gamma^{(a_2)}$ we write
\ben \label{eks3}
&& \kappa(\alpha_{n+1},
\gamma^{(a_1)}) e_{\gamma^{(a_1)}+ \alpha_{n+1}}
e_{\gamma^{(a_2)}} e_{\gamma^{(a_3)}} \cdots e_{\gamma^{(a_k)}}\nn
&=&
\kappa(\alpha_{n+1},
\gamma^{(a_1)}) e_{\gamma^{(a_2)}}
e_{\gamma^{(a_1)}+ \alpha_{n+1}}  e_{\gamma^{(a_3)}} \cdots e_{\gamma^{(a_k)}}
\nn && + \kappa(\alpha_{n+1},
\gamma^{(a_1)})
\kappa(\alpha_{n+1} + \gamma^{(a_1)},
\gamma^{(a_2)}) e_{\gamma^{(a_1)}+\gamma^{(a_2)}+ \alpha_{n+1}} 
 e_{\gamma^{(a_3)}} \cdots e_{\gamma^{(a_k)}}\, .
\een
In the next step we shall need to manipulate the product of 
$e_{\gamma^{(a_3)}}$  with the terms to its
left and so on.

To extract $g_{\rm KS}(\alpha_1,\cdots \alpha_{n+1})$
from this  we have to determine the coefficients
of $e_{\alpha_1+\cdots \alpha_{n+1}}$. Now by examining \refb{eks1} we
can see that the first term on the right hand side
can never contribute to this sum.
This is because we have $\gamma^{(a_1)}<\alpha_{n+1}, \gamma^{(a_2)},\cdots \gamma^{(a_k)}$.
Thus whatever manipulation we do to bring 
$e_{\gamma^{(a_1)}}
e_{\alpha_{n+1}} e_{\gamma^{(a_2)}} \cdots e_{\gamma^{(a_n)}}$ in the
clockwise order, the $e_{\gamma^{(a_1)}}$ will never be involved in the
manipulation and continue to sit at the left. Thus every term that we get from this
will have an $e_{\gamma^{(a_1)}}$ factor at the extreme left and we shall never
get $e_{\alpha_1+\cdots \alpha_{n+1}}$. By the same logic, the first term on
the right hand side of \refb{eks3} will never produce $e_{\alpha_1+\cdots \alpha_{n+1}}$.
 By repeated use of this logic we see that the only
term in $e_{\alpha_{n+1}} e_{\gamma^{(a_1)}} e_{\gamma^{(a_2)}} 
\cdots e_{\gamma^{(a_k)}}$ proportional to $e_{\alpha_1+\cdots \alpha_{n+1}}$
is given by
\be \label{eks4}
\kappa(\alpha_{n+1},
\gamma^{(a_1)})
\kappa( \alpha_{n+1} + \gamma^{(a_1)},
\gamma^{(a_2)}) \cdots 
\kappa(\alpha_{n+1} + \gamma^{(a_1)} +\cdots \gamma^{(a_{k-1})},
\gamma^{(a_k)}) \, e_{\alpha_{n+1} +\gamma^{(a_1)} +\cdots \gamma^{(a_k)}}\, ,
\ee
and furthermore this term exists only under the condition
\be \label{eks5}
\alpha_{n+1}+\gamma^{(a_1)} > \gamma^{(a_2)}, \quad
\alpha_{n+1}+\gamma^{(a_1)}+ \gamma^{(a_2)} > \gamma^{(a_3)}, \quad
\cdots, \quad \alpha_{n+1} + \gamma^{(a_1)} +\cdots \gamma^{(a_{k-1})}
> \gamma^{(a_k)}\, .
\ee
Using \refb{eksreversenew} we now get
\ben \label{eks6}
&& g_{\rm KS}(\alpha_1,\cdots \alpha_{n+1}) =
\sum_{k=1}^n
\sum_{\{a_1,\cdots a_k\}
\atop
\gamma^{(a_1)}+\cdots \gamma^{(a_k)}=\alpha_1+\cdots 
+\alpha_n; \, a_1<a_2\cdots <a_k}
\left(\prod_{\ell=1}^k g_{\rm KS}(\{\alpha_i, \, i\in A_{a_\ell}\}) \right)
\nn && \quad \times 
\Theta(\alpha_{n+1} + \gamma^{(a_1)},
\gamma^{(a_2)}) \Theta(\alpha_{n+1} + \gamma^{(a_1)}+
\gamma^{(a_2)}, \gamma^{(a_3)})\cdots 
\Theta(\alpha_{n+1} + \gamma^{(a_1)} +\cdots \gamma^{(a_{k-1)}},
\gamma^{(a_k)})\nn
&& \quad \times \kappa( \alpha_{n+1},
\gamma^{(a_1)})
\kappa(\alpha_{n+1} + \gamma^{(a_1)},
\gamma^{(a_2)}) \cdots 
\kappa(\alpha_{n+1} + \gamma^{(a_1)} +\cdots \gamma^{(a_{k-1)}},
\gamma^{(a_k)})\, , 
\een
where $\Theta(\gamma_1, \gamma_2)$ has been defined in \refb{enotations}.

\sectiono{Equivalence of KS and `higgs branch' wall crossing formul\ae} \label{s5}

We shall now prove the equivalence of $g_{\rm KS}$ and $g_{\rm higgs}$
using the method of induction, \i.e.\
we shall assume that $g_{\rm KS}(\alpha_1,\cdots \alpha_m)
= g_{\rm higgs}(\alpha_1,\cdots \alpha_m)$ for $m\le n$ and then prove the
result for $m=n+1$.
The equality of $g_{\rm KS}(\alpha_1, \alpha_2)$ and 
$g_{\rm higgs}(\alpha_1, \alpha_2)$ will then imply the equivalence of
$g_{\rm KS}(\alpha_1, \cdots \alpha_n)$ and 
$g_{\rm higgs}(\alpha_1, \cdots \alpha_n)$ for all $n$.

\subsection{$g_{\rm KS}$ as a sum over permutations} \label{s5.1}

Assuming the equality of $g_{\rm KS}(\alpha_1,\cdots \alpha_m)
= g_{\rm higgs}(\alpha_1,\cdots \alpha_m)$ for $m\le n$ we can replace $g_{\rm KS}$
by $g_{\rm higgs}$ on the right hand side of \refb{eks6} and get
\ben \label{emps1}
&& g_{\rm KS}(\alpha_1,\cdots \alpha_{n+1}) =
\sum_{K=1}^n
\sum_{\{a_1,\cdots a_K\}
\atop
\gamma^{(a_1)}+\cdots \gamma^{(a_K)}=\alpha_1+\cdots 
+\alpha_n; \, a_1<a_2\cdots <a_K}
\left(\prod_{\ell=1}^K g_{\rm higgs}(\{\alpha_i, \, i\in A_{a_\ell}\}) \right)
\nn && \quad \times 
\Theta(\alpha_{n+1} + \gamma^{(a_1)},
\gamma^{(a_2)})  \Theta(\alpha_{n+1} + \gamma^{(a_1)}+
\gamma^{(a_2)}, \gamma^{(a_3)}) \cdots 
\Theta(\alpha_{n+1} + \gamma^{(a_1)} +\cdots \gamma^{(a_{K-1)}},
\gamma^{(a_K)})\nn
&& \quad \times \kappa( \alpha_{n+1},
\gamma^{(a_1)})
\kappa(\alpha_{n+1} + \gamma^{(a_1)},
\gamma^{(a_2)}) \cdots 
\kappa(\alpha_{n+1} + \gamma^{(a_1)} +\cdots \gamma^{(a_{K-1)}},
\gamma^{(a_K)})\, .
\een
Note that we have replaced the summation variable $k$ by $K$ since soon we
shall use the variable $k$ for other purposes. 
Let $n_a$ be the total number of elements in the set $A_a$ and let
$I^{(a)}_1,\cdots I^{(a)}_{n_a}$ be the elements of $A_a$, ordered so that
$I^{(a)}_{1}<I^{(a)}_{2}<\cdots I^{(a)}_{n_a}$. 
After substituting the expression for $g_{\rm higgs}$ given in \refb{higgsfnewest}
we get
\ben \label{emps2one}
&& g_{\rm KS}(\alpha_1,\cdots \alpha_{n+1}) =\sum_{K=1}^n
\sum_{\{a_1,\cdots a_K\}
\atop
\gamma^{(a_1)}+\cdots \gamma^{(a_K)}=\alpha_1+\cdots 
+\alpha_n; \, a_1<a_2\cdots <a_K} (-1)^{n-K} (y-y^{-1})^{K-n}
 \nn && 
\sum_{\bar\sigma\atop \bar\sigma(A_{a_k})=A_{a_k} \, \forall\, k}
\Bigg[ \Bigg\{\prod_{\ell=1}^K \prod_{k=1
\atop \bar\sigma\left(I^{(a_\ell)}_{k+1}\right)<\bar\sigma
\left(I^{(a_\ell)}_k\right)}^{n_{a_\ell}-1}
\Theta\bigg(\sum_{j=1}^k \alpha_{\bar\sigma(I^{(a_\ell)}_j)},
\sum_{j=1}^{n_{a_\ell}} \alpha_{\bar\sigma(I^{(a_\ell)}_j)}
\bigg)  \Bigg\}\nn &&
\Bigg\{\prod_{\ell=1}^K\prod_{k=1
\atop \bar\sigma\left(I^{(a_\ell)}_{k+1}\right)>\bar\sigma
\left(I^{(a_\ell)}_k\right)}^{n_{a_\ell}-1}
\Theta\bigg(
\sum_{j=1}^{n_{a_\ell}} \alpha_{\bar\sigma(I^{(a_\ell)}_j)},
\sum_{j=1}^k \alpha_{\bar\sigma(I^{(a_\ell)}_j)}
\bigg)  \Bigg\}
(-y)^{\sum_{k=1}^K \sum_{l',l\in A_{a_k},\, l<l' }
\alpha_{\bar\sigma(l)\bar\sigma(l')}} \nn
&& \quad (-1)^{\sum_{\ell=1}^K \sum_{k=1}^{n_{a_\ell}-1} 
\Theta\left(
\bar\sigma\left(I^{(a_\ell)}_{k}\right)-\bar\sigma
\left(I^{(a_\ell)}_{k+1}
\right)\right)}\Bigg] \times
\Bigg\{
\prod_{k=1}^{K-1} \Theta\left(\alpha_{n+1}+\gamma^{(a_1)}
+\cdots \gamma^{(a_k)},
\gamma^{(a_{k+1})}\right)\Bigg\}
 \nn
&& \prod_{\ell=1}^{K}
\left( (-y)^{\sum_{i\in A_{a_\ell}} \alpha_{(n+1)i} + \sum_{r=1}^{\ell-1}
\sum_{j\in A_{a_\ell}, i\in A_{a_r}} \alpha_{ij}}
- (-y)^{-\sum_{i\in A_{a_\ell}} \alpha_{(n+1)i} - \sum_{r=1}^{\ell-1}
\sum_{j\in A_{a_\ell}, i\in A_{a_r}} \alpha_{ij}}
\right)\nn
&& \times (y-y^{-1})^{-K} 
\een
where in the expression the sum over $\bar\sigma$ denotes 
sum over a restricted
set of permutations each of which permutes the elements of the 
set $A_{a_k}$ among themselves for every $k$. 
We shall now express the factor in the last but one line
of \refb{emps2one} as
\be \label{emps2two}
\prod_{\ell=1}^{K}
\left( (-y)^{ \sum_{i\in A_\ell} \alpha_{(n+1)i}+
\sum_{r=1}^{\ell-1}
\sum_{j\in A_{a_\ell}, i\in A_{a_r}} \alpha_{ij}}
- (-y)^{\sum_{i\in A_{a_\ell}} \alpha_{i(n+1)} +\sum_{r=1}^{\ell-1}
\sum_{i\in A_{a_\ell}, j\in A_{a_r}} \alpha_{ij}}
\right)
\ee
and expand this as a sum of $2^K$ terms. After
substituting this into \refb{emps2one} we get 
\ben \label{emps2}
&& g_{\rm KS}(\alpha_1,\cdots \alpha_{n+1})\nn
&=& (-1)^{n} 
(y-y^{-1})^{-n} 
\sum_{K=1}^n
\sum_{\{a_1,\cdots a_K\}
\atop
\gamma^{(a_1)}+\cdots \gamma^{(a_K)}=\alpha_1+\cdots 
+\alpha_n; \, a_1<a_2\cdots <a_K} (-1)^K \nn &&
\sum_{\bar\sigma\atop \bar\sigma(A_{a_k})=A_{a_k} \, \forall\, k}
\Bigg[ \Bigg\{\prod_{\ell=1}^K \prod_{k=1
\atop \bar\sigma\left(I^{(a_\ell)}_{k+1}\right)<\bar\sigma
\left(I^{(a_\ell)}_k\right)}^{n_{a_\ell}-1}
\Theta\bigg(\sum_{j=1}^k \alpha_{\bar\sigma(I^{(a_\ell)}_j)},
\sum_{j=1}^{n_{a_\ell}} \alpha_{\bar\sigma(I^{(a_\ell)}_j)}
\bigg)  \Bigg\}\nn &&
\Bigg\{\prod_{\ell=1}^K\prod_{k=1
\atop \bar\sigma\left(I^{(a_\ell)}_{k+1}\right)>\bar\sigma
\left(I^{(a_\ell)}_k\right)}^{n_{a_\ell}-1}
\Theta\bigg(
\sum_{j=1}^{n_{a_\ell}} \alpha_{\bar\sigma(I^{(a_\ell)}_j)},
\sum_{j=1}^k \alpha_{\bar\sigma(I^{(a_\ell)}_j)}
\bigg)  \Bigg\}
(-y)^{\sum_{k=1}^K \sum_{l',l\in A_{a_k},\, l<l'}
\alpha_{\bar\sigma(l)\bar\sigma(l')}} \nn
&& (-1)^{\sum_{\ell=1}^K \sum_{k=1}^{n_{a_\ell}-1} \Theta\left(
\bar\sigma\left(I^{(a_\ell)}_{k}\right)-\bar\sigma
\left(I^{(a_\ell)}_{k+1}
\right)\right)}\Bigg]
\times  
\prod_{k=1}^{K-1} \Theta\left(\alpha_{n+1}+\gamma^{(a_1)}+\cdots \gamma^{(a_k)},
\gamma^{(a_{k+1})}\right) \, 
\nn
&& \sum_{q=0}^K \, (-1)^{K-q} \,
\sum_{\{s_1,\cdots s_q\}\atop 1\le s_1< s_2\cdots < s_q\le K}\nn &&
(-y)^{ \sum_{\ell \neq s_1, \cdots s_q} 
\left(\sum_{i\in A_{a_\ell}} \alpha_{i(n+1)} +\sum_{r=1}^{\ell-1}
\sum_{j\in A_{a_r}, i\in A_{a_\ell}} \alpha_{ij}\right)
+ \sum_{\ell= s_1, \cdots s_q} \left(\sum_{i\in A_{a_\ell}} \alpha_{(n+1)i}+
\sum_{r=1}^{\ell-1}
\sum_{i\in A_{a_r}, j\in A_{a_\ell}} \alpha_{ij} \right)}
\nn
\een
Here $s_1,\cdots s_q$ are the values of $\ell$ in \refb{emps2two} for which we
pick the first term of the factor, where for the rest of the values of
$\ell$ we pick the second factor.
The prefactor of $(-1)^{n}
(y-y^{-1})^{-n} $ on the right hand side of \refb{emps2}
matches a similar factor in \refb{higgsfnewest}
with $n$ replaced by $n+1$. Leaving aside these factors 
the net power of $(-y)$ in a given term in the right hand side of
\refb{emps2} is given by
\ben \label{emps3}
&&  \sum_{k=1}^K \, \, \sum_{l',l\in A_{a_k}\atop l<l'}
\alpha_{\bar\sigma(l)\bar\sigma(l')}
+ \sum_{\ell=1\atop \ell \neq s_1, \cdots s_q}^K
\sum_{i\in A_{a_\ell}} \alpha_{i(n+1)} 
+ \sum_{\ell=s_1, \cdots s_q}
\sum_{i\in A_{a_\ell}} \alpha_{(n+1)i}\nn &&
+\sum_{\ell=1\atop \ell \neq s_1, \cdots s_q}^K \sum_{r=1}^{\ell-1}
\sum_{j\in A_{a_r}, i\in A_{a_\ell}} \alpha_{ij} 
+ \sum_{\ell= s_1, \cdots s_q} \sum_{r=1}^{\ell-1}
\sum_{i\in A_{a_r}, j\in A_{a_\ell}} \alpha_{ij}
\een
This can be expressed as
\be \label{expre1}
\sum_{i,j=1\atop i<j}^{n+1} \alpha_{\sigma(i)\sigma(j)}\, ,
\ee
where $\sigma$ denotes a permutation of $\{1,\cdots n+1\}$ given by
\be \label{emps4}
\{{\sigma(1)}, \cdots {\sigma(n+1)}\}
= \{\bar\sigma(A_{a_K}), \cdots \bar\sigma(\not \hskip -4pt A_{a_{s_q}}), 
\cdots \bar\sigma(\not  \hskip -4pt
A_{a_{s_1}}) \cdots \bar\sigma(A_{a_1}), 
n+1 , \bar\sigma(A_{a_{s_1}}), 
\cdots \bar\sigma(A_{a_{s_q}})\}\, .
\ee
$\bar\sigma(A_{a_\ell})$ 
contains the elements of $A_{a_\ell}$ ordered according
to the permutation $\bar\sigma$ restricted to the set $A_{a_\ell}$. 
The symbol $\bar\sigma(\not \hskip -5pt A_{{a_\ell}})$ denotes that the
corresponding $\bar\sigma(A_{a_\ell})$ is missing from the list, since it
is placed on the right hand side of $n+1$.

We shall now try to
reorganize the sum in \refb{emps2} by first
summing over all contributions corresponding to a given permutation
$\sigma$ of $(1,\cdots n+1)$, and then summing over $\sigma$.
Let $R$ denote the position of $n+1$ on the right hand side
of \refb{emps4}, \i.e.\ $\sigma(R)=n+1$. We
introduce sets of integers
\be \label{edefbpart}
B_1=\{1,\cdots I_1-1\}, \quad B_2=\{I_1, \cdots I_2-1\}, \quad \cdots
\quad B_p = \{I_{p-1}, \cdots (R-1)\}\, ,
\ee
and
\be \label{edefcpart}
C_q =\{(R+1), \cdots J_1-1\}, \quad C_{q-1} = \{J_1, \cdots J_2-1\}, \quad \cdots \quad
C_1 = \{ J_{q-1}, \cdots (n+1)\}\, .
\ee
for appropriate integers $I_1,\cdots I_{p-1}$ and $J_1,\cdots J_{q-1}$
with $p=K-q$ and $I_1<I_2<\cdots I_{p-1}<R<J_1<J_2<\cdots <J_{q-1}$
such that 
\ben \label{econd}
&&\{ \{\sigma(B_1)\}, \cdots \{\sigma(B_p)\}\}
= \{\{\bar\sigma(A_{a_K})\}, \cdots 
\{\bar\sigma(\not \hskip -4pt A_{a_{s_q}})\}, 
\cdots \{\bar\sigma(\not  \hskip -4pt
A_{a_{s_1}})\} \cdots \{\bar\sigma(A_{a_1})\}\}\nn
&& \{\{\sigma(C_q)\},
\cdots \{\sigma(C_1)\}\}
=
\{\{\bar\sigma(A_{a_{s_1}})\}, 
\cdots \{\bar\sigma(A_{a_{s_q}})\}\}\, ,
\een
as ordered sets.
Thus $\sigma(B_a)$'s correspond to the sets of $\bar\sigma(A_{a_i})$'s
in \refb{emps4} to the left of $(n+1)$ and 
$\sigma(C_a)$'s correspond to the sets of $\bar\sigma(A_{a_i})$'s
in \refb{emps4} to the right of $(n+1)$.
We also define:\footnote{We shall use the same index $a$ for 
$B_a$, $C_a$, $\delta^{(a)}$ and
$\tau^{(a)}$. 
However it should be understood that for
$B_a$ and $\delta^{(a)}$ the index runs from 1 to $p$ and
for $C_a$ and $\tau^{(a)}$ the index runs from 1 to $q$.}
\ben \label{edefdeleps}
\delta^{(a)} &=& \sum_{i\in B_a} \alpha_{\sigma(i)}=
\cases{\sum_{i=1}^{I_1-1}\alpha_{\sigma(i)} \quad \hbox{for $a=1$}\cr
\sum_{i=I_{a-1}}^{I_a-1} \alpha_{\sigma(i)}\quad \hbox{for $
p-1\ge a\ge 2$}\cr
\sum_{i=I_{p-1}}^{R-1} \alpha_{\sigma(i)}\quad \hbox{for $a=p$}
}, \nn
\tau^{(a)} &=& \sum_{i\in C_a} \alpha_{\sigma(i)}
= \cases{\sum_{i=R+1}^{J_{1}-1} \alpha_{\sigma(i)}\quad 
\hbox{for $a=q$}\cr
\sum_{i=J_{q-a}}^{J_{q+1-a}-1} \alpha_{\sigma(i)}\quad 
\hbox{for $q-1\ge a\ge 2$}\cr
\sum_{i=J_{q-1}}^{n+1} \alpha_{\sigma(i)} \quad \hbox{for $a=1$}}\, .
\een
The $\delta^{(a)}$'s correspond to the $\gamma^{(a_i)}$'s for
$i=K, K-1,\cdots, \not \hskip -2pt s_q, \cdots, \not \hskip -2pt s_1,\cdots 1$
and $\tau^{(a)}$'s correspond to the
$\gamma^{(a_i)}$'s for $i=s_q, s_{q-1},\cdots , s_1$ 
in \refb{emps2}. A pictorial representation of this arrangement can be
given as follows:
{\small
\ben \label{epicto}
{B_1}  \hskip .7in \cdots \hskip .6in {B_p}  \hskip .9in {R} \hskip .9in {C_q}  
\hskip .75in \cdots \hskip .7in {C_1} \hskip .6in
\nonumber \\
{\sigma\downarrow}  \hskip 1.5in {\sigma\downarrow}  \hskip .9in {\sigma\downarrow} 
\hskip .8in {\sigma\downarrow}  \hskip 1.6in {\sigma\downarrow} \hskip .6in
\nonumber \\
{\{\sigma(1), \cdots \sigma(I_1-1)\}},
\cdots
{\{\sigma(I_{p-1}), \cdots \sigma(R-1)\}},
{\sigma(R)}, {\{\sigma(R+1), \cdots \sigma(J_1-1)\}},
\cdots {\{\sigma(J_{q-1}), \cdots \sigma(n+1)\}}
\nonumber \\
\delta^{(1)}  \hskip .7in \cdots \hskip .6in \delta^{(p)}  \hskip .7in \alpha_{n+1} \hskip .8in \tau^{(q)}  \hskip .7in \cdots \hskip .6in \tau^{(1)} \hskip .6in
\nonumber \\  
\een
}
\hskip -.09in
The last row describes the sum of all the $\alpha_{\sigma(i)}$'s in the sets
in the last but one row.
The partitioning described above  must satisfy the following constraints:
\begin{enumerate}
\item 
The restrictions
$a_1<a_2< \cdots a_K$ and $s_1<s_2<\cdots s_q$ in \refb{emps2}
translate to the following restrictions on $\{\delta^{(a)}\}$, $\{\tau^{(a)}\}$:
\be \label{emust}
\delta^{(1)} > \delta^{(2)} > \cdots > \delta^{(p)}, \quad 
\tau^{(1)} > \tau^{(2)} > \cdots > \tau^{(q)}\, .
\ee
\item Let us denote by $\wh\gamma^{(k)}$ ($1\le k\le p+q$) the set of vectors
$\{\delta^{(1)}, \cdots \delta^{(p)}, \tau^{(1)}, \cdots \tau^{(q)}\}$ ordered
so that $\wh\gamma^{(1)}< \wh\gamma^{(2)}< \cdots \wh\gamma^{(p+q)}$. Thus
we have $\wh\gamma^{(k)}=\gamma^{(a_k)}$ and 
the third set of $\Theta$'s in \refb{emps2} imposes the
constraints:
\be \label{emust2}
\alpha_{n+1} + \wh\gamma^{(1)} +\cdots \wh\gamma^{(k)} > \wh\gamma^{(k+1)}
\quad \hbox{for} \quad k= 1, 2, \cdots p+q-1\, .
\ee
Using \refb{euseful} this is equivalent to the condition
\be \label{emust2alt}
 \alpha_{n+1}+ \wh\gamma^{(1)} >\alpha_{n+1}+ \wh\gamma^{(1)}
+\wh\gamma^{(2)} > \cdots > \alpha_{n+1}+ \wh\gamma^{(1)}
+\wh\gamma^{(2)}
+\cdots + \wh\gamma^{(p+q)}=\bar\alpha\, .
\ee
\item Since for $I_{a-1}\le k\le I_a-1$, $k\in B_a$, which is one of the $A_{a_i}$'s
appearing in  \refb{emps2}, 
the first and second
set of $\Theta$'s in \refb{emps2} impose the constraints:
\ben \label{emust3}
&&
\delta^{(a)} > \sum_{i=I_{a-1}}^{k}  \alpha_{\sigma(i)}\quad \hbox{for} \quad
\sigma(k+1)> \sigma(k), \quad
\delta^{(a)} < \sum_{i=I_{a-1}}^{k} \alpha_{\sigma(i)}\quad \hbox{for} \quad
\sigma(k+1)< \sigma(k), \nn
&& \qquad \qquad \qquad \qquad I_{a-1}\le k \le I_{a}-2, \quad 1\le a\le p, 
\quad I_0\equiv 1, \quad I_p\equiv R\, .
\een
\item Similarly since
for $J_{a-1}\le k\le J_a-1$, $k\in C_{q-a+1}$, which is one of the $A_a$'s
appearing in  \refb{emps2}, 
the first and second
set of $\Theta$'s in \refb{emps2} impose the constraints:
\ben \label{emust4}
&& \tau^{(q+1-a)} > \sum_{j=J_{a-1}}^{k}  \alpha_{\sigma(j)}\quad \hbox{for} \quad
\sigma(k+1)> \sigma(k),  \nn
&& \tau^{(q+1-a)} < \sum_{j=J_{a-1}}^{k}  \alpha_{\sigma(j)}\quad \hbox{for} \quad
\sigma(k+1)< \sigma(k), \nn
&&  J_{a-1}\le k \le J_{a}-2, \quad 1\le a\le q, \quad J_0\equiv R+1, \quad J_q
\equiv n+2\, .\nn
\een
Note that under the reversal of the permutation associated with $\sigma$ the
roles of $\tau^{(a)}$ and $\delta^{(a)}$ get interchanged. 
\end{enumerate}

We can now run the arguments in reverse to
find an algorithm for computing $g_{\rm KS}$ as a sum
over permutations and partitions. 
Given any permutation $\sigma$, we consider all
possible choices of the sets $B_a$ and $C_a$ 
encoded in the choice of the integers
$p,q,I_1, \cdots I_{p-1}, J_1, \cdots J_{q-1}$. It is easy to see that for a
given choice of $\sigma$ and the integers 
$p,q,I_1, \cdots I_{p-1}, J_1, \cdots J_{q-1}$, the summation variable
$K=p+q$ and permutations $\bar\sigma$ in \refb{emps2}
are completely fixed. We then need to verify if the corresponding $\{\delta^{(a)}\}$
and $\{\tau^{(a)}\}$ satisfy the four conditions mentioned above.
If they do then we shall call this choice of
$\{p, q, I_1, \cdots I_{p-1}, J_1, \cdots J_{q-1}\}$ an allowed partition. 
The net contribution for a given $\sigma$ is then obtained by summing over
all the allowed
partitions weighted by the factors which appear in \refb{emps2}.
With the help of \refb{expre1}, the
contribution to $g_{\rm KS}(\alpha_1,\cdots \alpha_{n+1})$ given in
\refb{emps2} may then be written as 
\be \label{egmsfor}
g_{\rm KS}(\alpha_1,\dots ,\alpha_{n+1})
=(-1)^{n}  \,
 \left(y-y^{-1}\right)^{-n}
 \sum_{\sigma} {\rm N}^{(n+1)}_{\rm KS}(\{\alpha_i\};\sigma) (-y)^{
 \sum_{l<k}
 \alpha_{\sigma(l)
 \sigma(k) }}\, ,
 \ee
 where 
\be \label{esumperm}
{\rm N}^{(n+1)}_{\rm KS}(\{\alpha_i\};\sigma) = \sum_{\hbox{allowed partitions}} 
(-1)^{q +\sum_{a=1}^p {k_a} 
+ \sum_{a=1}^q {l_a}} \, ,
\ee
\be \label{edefkala}
k_a \equiv \sum_{k\in B_a, \, k+1\in B_a} \Theta(\sigma(k) - \sigma(k+1)), 
\qquad l_a \equiv \sum_{k\in C_a, \, k+1\in C_a} 
\Theta(\sigma(k) - \sigma(k+1))\,.
\ee
Our goal will be to show that ${\rm N}^{(n+1)}_{\rm KS}(\{\alpha_i\};\sigma)$ defined this way agrees with
the coefficient ${\rm N}^{(n+1)}_{\rm higgs}(\{\alpha_i\};\sigma)$
given in \refb{higgsfnewest}.

\subsection{Deforming the $\alpha_i$'s} \label{s5.2}

As mentioned in \S\ref{s1}, we have taken the $\alpha_i$'s to be
generic so that they have finite length and
the angle between two vectors of the form
$\sum_{i\in S_1}\alpha_i$ and $\sum_{i\in S_2}\alpha_i$ for any pair of
non-overlapping sets $S_1$, $S_2$ is non-vanishing. 
We have also seen that neither the result for ${\rm N}^{(n+1)}_{\rm KS}(\{\alpha_i\};\sigma)$
nor the result for ${\rm N}^{(n+1)}_{\rm higgs}(\{\alpha_i\};\sigma)$ changes under a deformation of the
$\alpha_i$'s 
which preserves the relative orientation of the
$\alpha_i$'s and the relative orientation of $\sum_{i=1}^k \alpha_{\sigma(i)}$
with respect to $\sum_{i=1}^{n+1}\alpha_i$ for all $k$.
Our strategy now
will be to use the freedom to deform the $\alpha_i$'s and by this process bring 
some of the angles and lengths arbitrarily close to zero -- much
smaller than the angles and lengths in the starting configuration.
Since  the angles
and lengths which we have brought arbitrarily close to zero are
now much smaller than the other angles and lengths which we do not change
during the deformation -- which we shall refer to as {\it generic}
lengths and angles -- the computation of ${\rm N}^{(n+1)}_{\rm KS}(\{\alpha_i\};\sigma)$ 
given in \refb{esumperm} will simplify
in this new configuration. We shall then compare this
${\rm N}^{(n+1)}_{\rm KS}(\{\alpha_i\};\sigma)$ 
to ${\rm N}^{(n+1)}_{\rm higgs}(\{\alpha_i\};\sigma)$ obtained by replacing
$n$ by $n+1$ in \refb{higgsfnewest}.

For notational simplicity it will be convenient
to define the following quantities associated with a given permutation
$\sigma$:
\be \label{enewdef}
\wt\alpha_i = \alpha_{\sigma(i)}, \qquad \bar\alpha = \sum_{j=1}^{n+1}\alpha_j
= \sum_{j=1}^{n+1}\wt\alpha_j, \qquad
\beta_k=\sum_{i=1}^k \wt\alpha_i, \qquad \hbox{for $1\le i,k\le n+1$}\, .
\ee
In this notation we have
the ordering
\be \label{ealorder}
\wt\alpha_i > \wt\alpha_j \quad \hbox{if $\sigma(i)>\sigma(j)$}
\quad \Rightarrow \quad \Theta(\wt\alpha_i,\wt\alpha_j)
=\Theta(\sigma(i)-\sigma(j))\, .
\ee
The allowed deformations are those which preserve the relative
ordering of the $\wt\alpha_i$'s and the relative ordering between
$\beta_k$ and $\bar\alpha$ for each $k$.

Now if we deform all the $\wt\alpha_i$'s at once, or even a pair of
$\wt\alpha_i$'s which are not placed next to each other in the
chain $\{\wt\alpha_1,\cdots \wt\alpha_{n+1}\}$, it will change
many $\beta_i$'s at once, and we need to ensure that none of these
$\beta_i$'s cross over from one side of $\bar\alpha$ to the other
side.
For this reason we shall deform the $\alpha_i$'s in nearest neighbor
pairs: take a pair $(\wt\alpha_j, \wt\alpha_{j+1})$ and deform it to
\be \label{edeco1}
(\wt\alpha_j +\lambda \wt\alpha_j, \wt\alpha_{j+1}+\lambda' \wt\alpha_{j+1})\, ,
\ee
with $(\lambda,\lambda')$ a pair of real numbers satisfying the following
conditions:
\begin{enumerate}
\item $\lambda, \lambda'>-1$.
\item $\lambda \wt\alpha_j +\lambda' \wt\alpha_{j+1} \propto \bar\alpha$.
This condition determines $\lambda'$ in terms of $\lambda$ and makes
this into a one parameter deformation.
\item At least one of $\lambda$ or $\lambda'$ is negative. We can for
definiteness take $\lambda$ to be negative.
\end{enumerate} 
Clearly \refb{edeco1} and 
the first condition above
ensures that in the new configuration $\wt\alpha_j$, $\wt\alpha_{j+1}$
preserve their directions. The first and the second condition ensure
that with the new $\wt\alpha_i$'s, the new $\bar\alpha$ remains parallel to the
original $\bar\alpha$ and continues to be directed along the first
quadrant. Finally the third condition ensures that at least one of the
vectors among $\wt\alpha_j$, $\wt\alpha_{j+1}$
reduces its length during this deformation. We have taken this
to be the vector $\wt\alpha_j$.

Now during this deformation all the $\beta_k$'s for $k<j$ are preserved,
while we add a vector 
$\lambda \wt\alpha_j +\lambda' \wt\alpha_{j+1} \propto \bar\alpha$
to the $\beta_k$'s
for $k\ge j+1$. Thus the $\beta_k$'s for $k<j$ and $k>j$ cannot
cross $\bar\alpha$, and as long as $\beta_j$ does not cross the
vector $\bar\alpha$, this map
preserves ${\rm N}^{(n+1)}_{\rm KS}$ and ${\rm N}^{(n+1)}_{\rm higgs}$.
We can increase the magnitude of the deformation till one of the following
situation is encountered:\footnote{Since our initial choice of vectors were generic 
we need not consider the case where two of these
events occur simultaneously except when it is occurs as a result
of some identity that holds for generic $\{\alpha_i\}$'s.}
\begin{enumerate}
\item The orientation of $\beta_j$ may approach that of
$\bar\alpha$.
In order to preserve $N^{(n+1)}_{\rm KS}$ and $N^{(n+1)}_{\rm higgs}$,
we must stop
the deformation infinitesimally before $\beta_j$ becomes exactly parallel
to $\bar\alpha$.
In this case we shall say that $\beta_j$
has become {\it almost parallel} to $\bar\alpha$. 
\item $\wt\alpha_j$ may approach zero. In this case 
$\beta_j=\beta_{j-1}+\wt\alpha_j\to\beta_{j-1}$. Thus
such a situation can arise before we encounter the first possibility
only if $\beta_j$ and $\beta_{j-1}$ were on the same side of
$\bar\alpha$  to begin with.
As before we need to stop the deformation infinitesimally before $\wt\alpha_j$
becomes exactly zero.
In this case we shall say that $\wt\alpha_{j}$ has been made {\it almost zero}.
\item
$\wt\alpha_{j+1}$ may approach zero. In this case $\beta_{j}
=\beta_{j+1}-\wt\alpha_{j+1}
\to\beta_{j+1}$ and hence such a situation can arise before encountering
the first case only if 
$\beta_j$ and $\beta_{j+1}$ were on the same side of
$\bar\alpha$ to begin with.
In this case we shall say that $\wt\alpha_{j+1}$ has been made almost zero.
\end{enumerate}
If $\beta_j$ becomes almost parallel to $\bar\alpha$ first we stop the
process here. Otherwise we can continue the process as follows.
If we have made $\wt\alpha_j$
almost zero then we can repeat the process with the pair
$(\wt\alpha_{j-1}, \wt\alpha_{j+1})$. The deformation will now affect both
$\beta_{j-1}$ and $\beta_j$, but $\beta_j=\beta_{j-1}+\wt\alpha_j$ 
is now almost equal to
$\beta_{j-1}$, and as long as we ensure that the deformation does
not take $\beta_{j-1}$ across $\bar\alpha$, 
$\beta_{j}$ also does not cross $\bar\alpha$. 
Similarly if
$\wt\alpha_{j+1}$ has been made almost zero, we can continue the
analysis with the pair $(\wt\alpha_{j}$, $\wt\alpha_{j+2})$.
Repeating this procedure we see that 
at any stage we work with a pair $(\wt\alpha_k, \wt\alpha_\ell)$ ($k\le j
<\ell$)
with all the intervening $\wt\alpha_i$'s zero. This process stops when
$\beta_k$ becomes almost parallel to $\bar\alpha$. Once this happens,
all the other $\beta_i$'s for $k< i<\ell$ (including the $\beta_j$
associated with the starting position) also become almost parallel
to $\bar\alpha$ since the corresponding $\wt\alpha_i$'s have already been
made almost zero. We note furthermore that 
by our previous argument (points 2 and 3 above)
the chain cannot continue
past a point $k_0$ for which $\beta_{k_0-1}$ and $\beta_{k_0}$ are
on the opposite sides of $\bar\alpha$. The situation can be represented as
\be \label{eposition}
\pmatrix{\hbox{position $i$} & & k & k+1 &k+2 & \cdots & \ell-2 & \ell-1 & \ell\cr\cr
\wt\alpha_i  &\simeq& \cdot & 0 & 0 & \cdots & 0 & 0 & \cdot \cr\cr
 \beta_i &\wr\hskip -3pt\parallel & \bar\alpha & \bar\alpha 
 & \bar\alpha &\cdots & \bar\alpha & \bar\alpha & \cdot}\, ,
 \ee
with all the $\beta_i$'s for $k\le i<\ell$ being on the same side
of $\bar\alpha$. 
Here the symbol $\wr\hskip -3pt\parallel$ denotes `almost parallel to'.
The only exception to this is when in the above diagram $k=1$ or
$\ell-1=n+1$. Such a situation can arise if in the starting configuration
all the $\beta_i$'s for $i\le j$ were on the same side of $\bar\alpha$
or all the $\beta_i$'s for $i\ge j$ were on the same side of
$\bar\alpha$. 
In the former case we may arrive at a configuration
in which all the $\wt\alpha_i$'s and
$\beta_i$'s for $1\le i < \ell$ are almost
zero but none of the $\beta_i$'s are almost parallel to $\bar\alpha$.
In the latter case we can arrive at a situation where 
all the $\wt\alpha_i$'s for $k<i\le n+1$ are almost
zero and all the $\beta_i$'s  for $k\le i\le n$ are
almost  equal (and hence almost parallel) to $\bar\alpha$
since by definition $\beta_{n+1}=\bar\alpha$.

In what follows, the neighborhood of the location of $\alpha_{n+1}$ will
play a special role. 
We shall denote the position of $\alpha_{n+1}$ by $R$, \i.e.\
$\wt\alpha_R=\alpha_{n+1}$. 
Thus $R$ marks the
maximum of $\sigma(i)$.
We shall carry out the manipulation described above by taking 
our starting pair to be $(R-1, R)$. Except for the special cases mentioned in the
last paragraph, which will be discussed separately later, 
at the end of the manipulation
we shall arrive at a situation where $\beta_{R-1}$ and possibly some other
$\beta_i$'s around it have been made almost parallel to $\bar\alpha$,
and some of the $\wt\alpha_i$'s around $R$
have been made almost zero.
If the set of points where $\wt\alpha_i$ becomes almost zero
includes also the point $R$, then we do not carry out any further
deformation of this system. If on the other hand it does not extend
beyond $R-1$ ({\it e.g.} for the case when $\beta_{R-1}$ and
$\beta_R$ are on the opposite sides of $\bar\alpha$)
then we start with the pair $(R, R+1)$ and carry out a
similar manipulation. At the end of this process we 
can have several situations:
\begin{enumerate}
\item Generically we
would get a
configuration in which we have
\be \label{efincon}
\pmatrix{\hbox{position} & & P-1 & P & P+1 & P+2 & \cdots &R-1 & R & \cdots & Q-2 & Q-1 & Q\cr\cr
\wt\alpha_i  &\simeq& \cdot & \cdot & 0 & 0 & \cdots & 0 & 0 &\cdots  & 0 & 0 & 
\cdot
 \cr\cr
 \beta_i &\wr\hskip -3pt\parallel & \cdot & \bar\alpha & \bar\alpha 
 & \bar\alpha &\cdots & \bar\alpha & \bar\alpha &\cdots & \bar\alpha & \bar\alpha 
 & \cdot}\, ,
 \ee
 for some positions $R>P>1$ and $R<Q<n+1$ in the chain. Furthermore 
 all the $\beta_i$'s for $P\le i\le R-1$ must be on the same side of
 $\bar\alpha$ and all the $\beta_i$'s for $R\le i\le Q-1$ 
 must be on the same side of
 $\bar\alpha$. Whether these two sets of $\beta_i$'s lie on the
 same side of $\bar\alpha$ or not depends on whether in the initial
 configuration $\beta_{R-1}$ and $\beta_R$ lie on the same side 
 of $\bar\alpha$ or not. 
 \item If in the starting configuration all the $\beta_i$'s for $i\le R-1$ are on the
same side of $\bar\alpha$ then the chain may continue all the way to the
left, setting all the $\wt\alpha_i$'s for $i\le R-1$ to almost zero. 
Similarly if
all the $\beta_i$'s for $i\ge R$ are on the same side of $\bar\alpha$ then
the chain may continue all the way to the right setting $\wt\alpha_i$ to be
almost zero for all $i> R$. We shall consider these special cases separately.
\end{enumerate}

 It is useful to note that under a reversal of permutation,
 the role of $\beta$ is played by $\sum_{i=j}^{n+1}\wt\alpha_i
 = \bar\alpha-\beta_{j-1}$.
 Thus to see the role of reversing the permutation we can express 
 \refb{efincon}  as
 \be \label{efinconrev}
\pmatrix{\hbox{position} & & Q & Q-1 & Q-2 & Q-3 & \cdots &R & R-1 & \cdots & P+1& P & P-1\cr\cr
\wt\alpha_i  &\simeq& \cdot & 0 & 0 & 0 & \cdots & 0 & 0 &\cdots  & 0 & \cdot & 
\cdot
 \cr\cr
 \bar\alpha - \beta_{i-1} &\wr\hskip -3pt\parallel & \bar\alpha & \bar\alpha & \bar\alpha 
 & \bar\alpha &\cdots & \bar\alpha & \bar\alpha &\cdots & \bar\alpha & \cdot
 & \cdot}\, .
 \ee
 Comparing \refb{efincon} and \refb{efinconrev} we see that the roles of
 the points $P$ and $Q$ are exchanged under a reversal of permutation.

\subsection{Constraining the permutations and partitions} \label{econper}

We shall now show that by making use of the deformations described in
\S\ref{s5.2} we can put severe restrictions on the permutations $\sigma$,
as well as the choices of the sets $\{B_a\}$, $\{C_a\}$, which
can contribute to ${\rm N}^{(n+1)}_{\rm KS}(\{\alpha_i\};\sigma)$.
In particular we shall show that 
\begin{enumerate}
\item ${\rm N}^{(n+1)}_{\rm KS}(\{\alpha_i\};\sigma)$ vanishes unless
\be \label{ebetapair}
\beta_{R-1}< \bar\alpha <\beta_R\, .
\ee
Combining this with the previous results we can also conclude that
we must have 
\be \label{eaddcond}
\hbox{$\beta_i<\bar\alpha$ for $P\le i\le R-1$, \qquad
$\beta_i>\bar\alpha$ for $R\le i<Q$}\, .
\ee 
\item For a configuration satisfying \refb{ebetapair} the choice of
the sets $B_1$ and $C_1$ 
described in \refb{edefbpart}, \refb{edefcpart} 
must be such that
\be \label{econsi1j}
I_1\ge P+1, \quad J_{q-1}\le Q\, .
\ee
Thus  $B_1$ should include at least the elements $(1,2,\cdots P)$ and
$C_1$ should include at least the elements $(Q,Q+1, \cdots n+1)$.
\item
In order that a permutation contributes
to ${\rm N}^{(n+1)}_{\rm KS}$ we must have
\be \label{econse2}
\beta_{k} \cases{\hbox{$< \bar\alpha$ for 
$\sigma(k+1)>\sigma(k)$, }\cr
\hbox{$>\bar\alpha$ for $\sigma(k+1)<\sigma(k)$},
}\, , \quad  1\le k < P  \, ,
\ee
\be \label{econse1}
\beta_{k} \cases{\hbox{$< \bar\alpha$ for 
$\sigma(k+1)>\sigma(k)$, }\cr
\hbox{$>\bar\alpha$ for $\sigma(k+1)<\sigma(k)$},
}\, , \quad Q\le k < n+1\, .
\ee
\end{enumerate}

\noindent {\bf Proof of \refb{ebetapair}}: 
If $\beta_{R-1}>\bar\alpha$ then clearly $\beta_{R}=
\beta_{R-1}+\wt\alpha_R=\beta_{R-1}+\alpha_{n+1}>\bar\alpha$. Thus
to prove \refb{ebetapair}
we only have to show that ${\rm N}^{(n+1)}_{\rm KS}$ vanishes
when $\beta_R, \beta_{R-1}>\bar\alpha$ or
$\beta_R, \beta_{R-1}<\bar\alpha$. These two cases
are in fact related by permutation reversal, so it is enough to
consider one of them. We shall consider the case 
$\beta_{R-1}, \beta_R <\bar\alpha$.

We proceed as in \S\ref{s5.2}, reducing $\wt\alpha_i$'s for
$i<R$ one by one by starting with the pair $(R-1, R)$.
As we have seen, when $\beta_P$ is almost parallel
to $\bar\alpha$ with
all the $\wt\alpha_i$ for $P<i\le R-1$ almost zero, then 
all $\beta_k$ for $P\le k\le R-1$ 
become almost parallel to $\bar\alpha$. In particular we have
$\beta_{R-1}\simeq c\bar\alpha$ for some positive constant $c$.
If at this stage $\wt\alpha_R=\alpha_{n+1}$ has finite length,
then we shall have $\beta_R=\beta_{R-1}+\wt\alpha_R\simeq
c\bar\alpha+\alpha_{n-1}>\bar\alpha$, contradicting our assumption that the
starting configuration has $\beta_R<\bar\alpha$. 
This shows that before we reach the stage where
$\beta_P$
becomes almost parallel to $\bar\alpha$, 
$\wt\alpha_R=\alpha_{n+1}$ should
become almost zero. 
Let us stop the
deformation as soon as $\wt\alpha_R$ becomes almost
zero and try to check if the required conditions are satisfied by any
choice of the sets $\{B_a\}$, $\{C_a\}$. 
Since we have stopped the deformation at a stage 
where some of the $\wt\alpha_i$'s may be almost zero
but none of the $\beta_i$'s are almost parallel to $\bar\alpha$,
at least $\delta^{(1)}$ and $\tau^{(1)}$, which can be identified as
$\beta_{I_1-1}$ and $\bar\alpha - \beta_{J_{q-1}-1}$, remain generic.
Now \refb{emust} shows that
when we order the $\delta^{(a)}$'s and $\tau^{(a)}$'s into the
sets $\wh\gamma^{(k)}$ with $\wh\gamma^{(1)}, \cdots \wh\gamma^{(p+q)}$
in the order $\wh\gamma^{(1)}<\wh\gamma^{(2)}< \cdots <\wh\gamma^{(p+q)}$,
then $\wh\gamma^{(p+q)}$ must be either $\delta^{(1)}$ or $\tau^{(1)}$
depending on whether $\delta^{(1)}>\tau^{(1)}$ or $\tau^{(1)}>\delta^{(1)}$. 
Thus $\wh\gamma^{(p+q)}$ and hence also $\bar\alpha-\wh\gamma^{(p+q)}
=\alpha_{n+1}+\wh\gamma^{(1)}+\wh\gamma^{(2)}+\cdots \wh\gamma^{(p+q-1)}$
are generic, \i.e.\
neither almost zero nor almost parallel to $\bar\alpha$.
In this case for $k=p+q-1$
we can drop the $\alpha_{n+1}$ from the left hand
side of \refb{emust2} since it has been made almost
zero, and express \refb{emust2} as
\be \label{eewmust}
\wh\gamma^{(1)}+\cdots \wh\gamma^{(p+q-1)} > \wh\gamma^{(p+q)}\, .
\ee
This is in obvious contradiction to the fact that $\wh\gamma^{(k)}$'s are
ordered as
\be \label{egamor}
\wh\gamma^{(1)}<\wh\gamma^{(2)}<\cdots < \wh\gamma^{(p+q)}\, .
\ee
Thus we see that there is no choice of the sets
$\{B_a\}$, $\{C_a\}$ satisfying the necessary conditions.
This shows that unless \refb{ebetapair} holds, ${\rm N}^{(n+1)}_{\rm KS}(\{\alpha_i\};\sigma)$
will vanish.

\noindent{\bf Proof of \refb{econsi1j}:}
We shall now examine, for a configuration satisfying
\refb{ebetapair},  possible ways
of dividing the set to the left of $R$ into the sets $\{B_a\}$ 
and the set to the right of $R$
into the sets $\{C_a\}$ subject to the 
constraints given in \refb{emust}-\refb{emust4}. 
For this we shall carry out the deformations all the
way to the end so that the final configuration at the
end of the deformation takes the form \refb{efincon}.
Now it follows from the ordering $\wh\gamma^{(1)}<\wh\gamma^{(2)}
< \cdots <\wh\gamma^{(p+q)}$ that for any $k$ we have
\be \label{ealt1}
\wh\gamma^{(1)}+\cdots \wh\gamma^{(k)} < \wh\gamma^{(k+1)}
+\wh\gamma^{(k+2)} + \cdots + \wh\gamma^{(p+q)}\, .
\ee
On the other hand, it follows from \refb{emust2alt} that
\be \label{e514alt}
\alpha_{n+1}+ \wh\gamma^{(1)}+\cdots \wh\gamma^{(k)} > 
\bar\alpha  - \left(\alpha_{n+1}+ \wh\gamma^{(1)}+\cdots \wh\gamma^{(k)}\right)
=\wh\gamma^{(k+1)}
+\wh\gamma^{(k+2)} +\cdots + \wh\gamma^{(p+q)}\, .
\ee
Let $k_0$ be the minimum value of $k$ for which $\wh\gamma^{(k_0)}$
is generic, \i.e.\ neither almost zero nor almost parallel to $\bar\alpha$.
In this case for $k=k_0$  we can drop the $\alpha_{n+1}$ from the left
hand side of \refb{e514alt} since $\alpha_{n+1}$ has been made almost
zero, and express it as
\be \label{ealt2}
\wh\gamma^{(1)}+\cdots \wh\gamma^{(k_0)} > \wh\gamma^{(k_0+1)}
+ \wh\gamma^{(k_0+2)} +\cdots + \wh\gamma^{(p+q)}\, .
\ee
This is in obvious contradiction to \refb{ealt1} for $k=k_0$
showing that our initial assumption must be wrong. In other words
all the $\wh\gamma^{(k)}$'s must be either almost zero or almost
parallel to $\bar\alpha$. Since the set $\{\wh\gamma^{(k)}\}$ includes
$\delta^{(1)}=\sum_{i=1}^{I_1-1}\wt\alpha_i
$ and $\tau^{(1)}=\sum_{i=J_{q-1}}^{n+1} \wt\alpha_i$, 
they must also satisfy this criterion.
This can happen only if 
$I_1>P$ and $J_{q-1}\le Q$ since otherwise $\delta^{(1)}$ 
and/or $\tau^{(1)}$ will involve
a sum of $\wt\alpha_i$'s which have not been deformed and hence
must be generic.
Thus
we must satisfy \refb{econsi1j}, and as a consequence $\delta^{(1)}$ and
$\tau^{(1)}$ are almost parallel to $\bar\alpha$.

\noindent{\bf Proof of \refb{econse2}, \refb{econse1}:}
For this we first test 
\refb{emust3} for $a=1$. 
Since $I_0=1$,  we have $\sum_{i=I_0}^{k} \alpha_{\sigma(i)}=\beta_k$
for $k\in B_1$. 
Since for $i<P$ we have not deformed $\wt\alpha_i$'s and $\beta_i$'s, 
they
are generic.
On the other hand as argued above $\delta^{(1)}$ appearing in
\refb{emust3} for $a=1$
is almost parallel to $\bar\alpha$.
Hence in testing \refb{emust3} for $a=1$ and $k < P$, replacing
$\delta^{(1)}$ by $\bar\alpha$ does not make any difference. 
After making these replacements we get $\beta_k<\bar\alpha$ for
$\sigma(k+1)>\sigma(k)$ and $\beta_k>\bar\alpha$ for
$\sigma(k+1)<\sigma(k)$ for $1\le k< P$. This gives \refb{econse2}.
Similarly testing \refb{emust4} for $a=q$ 
we get   \refb{econse1} since this is related to the previous case by
a reversal of permutation.

\subsection{Comparison with the 
constraints on $\sigma$ for non-vanishing ${\rm N}^{(n+1)}_{\rm higgs}(\{\alpha_i\};\sigma)$}
\label{econmps}

The constraints on $\sigma$ and the choice of the sets $\{B^{(a)}\}$,
$\{C^{(a)}\}$ derived in \S\ref{econper} are necessary but not sufficient
for getting a non-vanishing contribution to ${\rm N}^{(n+1)}_{\rm KS}(\{\alpha_i\};\sigma)$.
Nevertheless it will be useful at this stage to compare them with the
constraints on $\sigma$ needed for ${\rm N}^{(n+1)}_{\rm higgs}(\{\alpha_i\};\sigma)$ to be
non-vanishing.
According to \refb{higgsfnewest}
the latter constraints are given by:
\be \label{eallowed}
\beta_{k} \cases{\hbox{$< \bar\alpha$ for 
$\sigma(k+1)>\sigma(k)$, }\cr
\hbox{$>\bar\alpha$ for $\sigma(k+1)<\sigma(k)$},
}\, , \qquad \hbox{for $1\le k\le n$}\, .
\ee
Since $\sigma(R)=n+1$ is larger than both
$\sigma(R-1)$ and $\sigma(R+1)$, \refb{eallowed} for
$k=R-1$ and $k=R$ gives:
\be \label{eallowedfina}
\beta_{R-1} <\bar\alpha, \quad \beta_R>\bar\alpha\, .
\ee

The conditions on $\beta_{R-1},\beta_R$ given in \refb{eallowedfina}
agree with the corresponding constraints \refb{ebetapair} required for
${\rm N}^{(n+1)}_{\rm KS}$ to be non-vanishing. 
Furthermore, comparing \refb{eallowed} with 
\refb{econse2}, \refb{econse1} we see that in the range
$1\le k<P$ and $Q\le k\le n$, the condition on 
$(\beta_k, \wt\alpha_k, \wt\alpha_{k+1})$
needed 
for getting non-zero
${\rm N}^{(n+1)}_{\rm higgs}$ agrees with the 
condition on $(\beta_k, \wt\alpha_k, \wt\alpha_{k+1})$
needed for getting non-zero
${\rm N}^{(n+1)}_{\rm KS}$. Thus we need to focus on the
$(\beta_k, \wt\alpha_k, \wt\alpha_{k+1})$'s for 
$P\le k\le Q-1$ and show that for these
also the conditions agree. 
Since  all the $\beta_k$'s for
$P\le k\le R-1$ are on the same side of $\bar\alpha$ and all the
$\beta_k$'s for $R\le k<Q$ are on the same side of $\bar\alpha$,
we see from \refb{eallowedfina} that we have $\beta_k<\bar\alpha$
for $P\le k\le R-1$ and $\beta_k>\bar\alpha$ for $R\le k\le Q-1$.
Thus \refb{eallowed} takes a
simple form in the range $P\le k\le Q-1$:
\be \label{eallowedfin}
\hbox{$\sigma(k+1)>\sigma(k)$ for $P\le k\le R-1$}, \quad
\hbox{$\sigma(k+1)<\sigma(k)$ for $R\le k\le Q-1$}\, ,
\ee
\i.e.\ 
the $\wt\alpha_i$'s between $P$ and $R$ must be in the increasing
sequence and the $\wt\alpha_i$'s between $R$ and $Q$ 
must be in the decreasing
sequence.
Furthermore for configurations satisfying \refb{eallowedfin},
${\rm N}^{(n+1)}_{\rm higgs}$ given in \refb{higgsfnewest} takes the form
\ben \label{enmpspred}
{\rm N}^{(n+1)}_{\rm higgs}(\{\alpha_i\};\sigma) = (-1)^{s(\sigma)-1}\, ,
\quad\qquad\qquad\qquad\qquad\qquad\qquad\qquad
\qquad\qquad 
\nn 
s(\sigma)-1 = (Q-R) + 
\sum_{k=1}^{P-1} \Theta(\sigma(k)-\sigma(k+1)) 
+ \sum_{k=Q}^{n} \Theta(\sigma(k)-\sigma(k+1))\, ,
\een
where the additive factor of $Q-R$ arises from the contribution
for $R\le k\le Q-1$. 
Thus it remains to prove that for configurations satisfying \refb{eaddcond}, 
\refb{econse2}, \refb{econse1},
\begin{enumerate}
\item ${\rm N}^{(n+1)}_{\rm KS}$  is non-vanishing only for configurations which
satisfy \refb{eallowedfin}.
\item For these configurations ${\rm N}^{(n+1)}_{\rm KS}$ computed from \refb{esumperm}
agrees with ${\rm N}^{(n+1)}_{\rm higgs}$ given in  \refb{enmpspred}.
\end{enumerate}

\subsection{Proof of ${\rm N}^{(n+1)}_{\rm KS}={\rm N}^{(n+1)}_{\rm higgs}$} \label{s5.5}

We shall now compute ${\rm N}^{(n+1)}_{\rm KS}$ for a given permutation $\sigma$ and
show that the result agrees with that of ${\rm N}^{(n+1)}_{\rm higgs}$. 
We shall begin by analyzing the constraints on $\wt\alpha_k$,
$\beta_k$ for $k\le R-1$, \i.e.\ for points to the left of $R$.
By using the deformation we have set the $\wt\alpha_i$'s
in the range $P<k\le R-1$
to be almost zero but so far we have not said anything about their
relative magnitudes. We now note that the order in which the vectors
are reduced to almost zero during our manipulation is 
$(\wt\alpha_{R-1}, \wt \alpha_{R-2}, \cdots \wt\alpha_{P+1})$, \i.e.\
we first make $\wt\alpha_{R-1}$ almost zero, then $\wt \alpha_{R-2}$ almost
zero and so on. Thus we can arrange the deformations such that
the magnitudes of the $\wt\alpha_i$ are arranged in the order:
\be \label{emag}
|\wt\alpha_{R-1}|<< |\wt \alpha_{R-2}| << \cdots << |\wt\alpha_{P+1}|\, ,
\ee
where now the inequalities refer to standard inequalities between
ordinary numbers. 
This leads to the equation
\be \label{emag2}
\sum_{i=\ell}^{\ell'} \wt\alpha_i \simeq \wt\alpha_\ell \quad
\hbox{for $P<\ell\le \ell'\le R-1$}\, .
\ee
It now follows that
\be \label{emag3}
\delta^{(a)} = \sum_{i=I_{a-1}}^{I_a -1} \wt\alpha_i\simeq 
\wt\alpha_{I_{a-1}}\, , \quad \hbox{for $a\ge 2$}\,,
\ee
since we have shown earlier that $I_1>P$ and hence $I_{a-1}>P$ for
$a\ge 2$. 
The condition \refb{emust} and the fact that $\delta^{(1)}$ is almost parallel
to $\bar\alpha$  now gives
\be \label{emag4}
\bar\alpha > \wt\alpha_{I_1} > \wt\alpha_{I_2} > \cdots 
> \wt\alpha_{I_{p-1}}\, .
\ee
Let us now examine condition \refb{emust3} by expressing it as
\ben \label{emust3rep}
&&
\delta^{(a)} < \sum_{i=k+1}^{I_a-1}  \alpha_{\sigma(i)}\quad \hbox{for} \quad
\sigma(k+1)> \sigma(k), \quad 
\delta^{(a)} > \sum_{i=k+1}^{I_a-1} \alpha_{\sigma(i)}\quad \hbox{for} \quad
\sigma(k+1)< \sigma(k), \nn
&& \qquad \qquad \qquad \qquad I_{a-1}\le k \le I_{a}-2\, .
\een
Using $\wt\alpha_i\equiv \alpha_{\sigma(i)}$ and
\refb{emag2}, \refb{emag3} we can write this as
\ben \label{emag5}
&&\wt \alpha_{I_{a-1}} < \wt\alpha_{k+1}\quad \hbox{for} \quad
\wt\alpha_{k+1}> \wt\alpha_{k},
\quad \wt \alpha_{I_{a-1}} > \wt\alpha_{k+1}\quad \hbox{for} \quad
\wt\alpha_{k+1}< \wt\alpha_{k}, 
\nn
&& \qquad \qquad \qquad \qquad \hbox{for} \, \, I_{a-1}\le k \le I_{a}-2, \quad a\ge 2
\, , \nn
&& \bar\alpha < \wt\alpha_{k+1}\quad \hbox{for} \quad
\wt\alpha_{k+1}> \wt\alpha_{k},
\quad \bar\alpha > \wt\alpha_{k+1}\quad \hbox{for} \quad
\wt\alpha_{k+1}< \wt\alpha_{k}, 
\nn
&& \qquad \qquad \qquad \qquad \hbox{for} \, \, P\le k \le I_{1}-2\, .
\een
A similar set of constraints can be derived for the $\wt\alpha_i$'s for
$R<i<Q$ by working on the other side of $R$.

Finally we have to worry about the constraint \refb{emust2}. 
Since we have the $\beta_i$'s for $P\le i\le R-1$
and the $\beta_i$'s for $R\le i< Q$ on opposite sides of $\bar\alpha$,
it follows from our previous discussion that while manipulating the elements
on the left hand side of $R$, the elements on the right hand side of $R$
remain fixed. At the end of the first set of deformations $\wt\alpha_{n+1}$ 
as well as all the $\wt\alpha_i$'s on the right hand side of $R$ 
remain finite.
During the second set of deformations involving elements on the right hand
side of $R$ also $\wt\alpha_{n+1}$ remains finite almost till the end,
and becomes almost zero only at the very last stage. Thus its magnitude can be
taken to be larger than that of all the other almost zero $\wt\alpha_i$'s.
We can now consider three possible situations depending on the value we
take for $k$ in \refb{emust2}:
\begin{enumerate}
\item
The $\wh\gamma^{(i)}$'s which appear
in the sum on the left hand side of \refb{emust2} contains only 
$\delta^{(a)}$'s or $\tau^{(a)}$'s for $a\ge 2$. Since these are smaller
in magnitude compared to $\alpha_{n+1}$,
while testing  \refb{emust2}  we can replace the left hand side of this
equation by $\alpha_{n+1}$. 
In such cases these equations hold trivially since any linear combinations 
of the $\alpha_i$'s with non-negative coefficients is $<\alpha_{n+1}$. 
Let $\ell_0$ be the integer such that for all
$k<\ell_0$ the situation is as described above, \i.e.\ for all $k<\ell_0$,
$\wh\gamma^{(k)}$ corresponds to either $\delta^{(a)}$ or $\tau^{(a)}$
with $a\ge 2$.

\item For $k=\ell_0$ the $\wh\gamma^{(\ell_0)}$ in the sum is either $\delta^{(1)}$ or
$\tau^{(1)}$ depending on whether $\delta^{(1)}<\tau^{(1)}$ or 
$\tau^{(1)}<\delta^{(1)}$. Let us for definiteness assume that this is
$\tau^{(1)}$.
Now the left hand side of
\refb{emust2} will become almost equal to $\tau^{(1)}$
and hence is almost parallel to $\bar\alpha$. 
But now the $\wh\gamma^{(\ell_0+1)}$ on the right hand of the equation is either
$\delta^{(a)}$ or $\tau^{(a)}$ for some $a\ge 2$ and hence is,
by eqs.\refb{emag3}, \refb{emag4} and the corresponding equations for
$\tau^{(a)}$,
$<\bar\alpha$ and  not almost parallel
to $\bar\alpha$. 
Thus \refb{emust2} still holds.
The same argument holds for all subsequent values of $k$ till $k=p+q-2$, with the 
left hand
side almost parallel to $\bar\alpha$ and the right hand side $<\bar\alpha$ and not
almost parallel to $\bar\alpha$.

\item For $k=p+q-1$ the right hand side of \refb{emust2} becomes
$\wh\gamma^{(p+q)}=\delta^{(1)}$.
The left hand side of the equation is now $\bar\alpha -\delta^{(1)}$. Since both
sides are almost parallel to $\bar\alpha$ we need to carry out the comparison
with a little more care. For this note that \refb{emust2}, which requires
$\bar\alpha - \delta^{(1)}>\delta^{(1)}$ is equivalent to requiring 
$\delta^{(1)}<\bar\alpha$.
Since 
$\delta^{(1)}=\beta_{I_1-1}$ and
$P\le I_1-1\le R-1$, \refb{eaddcond} ensures that $\delta^{(1)}<\bar\alpha$.
Thus \refb{emust2} still holds.
\end{enumerate}
{}From this we conclude that \refb{emust2} does not impose any
additional constraints on the $\beta_k$'s and $\wt\alpha_k$'s once the
other conditions have been satisfied. A similar conclusion holds for the
$\beta_k$'s and $\wt\alpha_k$'s for $k>R$.

We now need to compute the contribution to ${\rm N}^{(n+1)}_{\rm KS}$ from the
allowed configurations.
The net contribution to ${\rm N}^{(n+1)}_{\rm KS}$ comes from summing the weight
factor given in \refb{esumperm} over all
possible choice of $p$ and $q$ and the integers 
$I_1,\cdots I_{p-1}$, $J_1, \cdots J_{q-1}$ 
subject to all the constraints. Now of the constraints given in 
\refb{emust}-\refb{emust4}, the constraints \refb{emust2} (or equivalently
\refb{emust2alt}) are the only ones which involve both the sets
$\{B_a\}$ and
$\{C_a\}$. Since we have argued that these constraints are
automatically satisfied when the other constraints are satisfied,
the constraints on $\{p, I_1,\cdots I_{p-1}\}$ and $\{q, J_1, \cdots J_{q-1}\}$
become independent of each other and hence we can carry out the sum
over $\{p, I_1,\cdots I_{p-1}\}$ and $\{q, J_1, \cdots J_{q-1}\}$
independently, and express \refb{esumperm} as
\be \label{e519}
\sum_{\hbox{\tiny allowed values of}\atop
\{p, I_1,\cdots I_{p-1}\}} (-1)^{\sum_{a=1}^p\sum_{k=I_{a-1}}^{I_a-1}
\Theta(\sigma(k)-\sigma(k+1))} \times
\sum_{\hbox{\tiny allowed values of}\atop\{q, J_1,\cdots J_{q-1}\}} (-1)^{q+
\sum_{a=1}^q\sum_{k=J_{a-1}}^{J_a-1}
\Theta(\sigma(k)-\sigma(k+1))}\, .
\ee
We shall first carry out the sum over $p$ and $I_1,\cdots I_{p-1}$.
Besides the constraints
given in \refb{emag4}, \refb{emag5}, we also need to account for the 
constraint \refb{econsi1j} that $I_1$ must be $\ge P+1$.
Taking into account all the constraints and introducing a new variable
$k=p-1$ we may express the net contribution as:
\ben \label{enetc}
&& (-1)^{\sum_{i=1}^{P-1} \Theta(\wt\alpha_i, \wt\alpha_{i+1})}\,
\Bigg\{ \prod_{i=P}^{R-2}
\bigg[ \Theta\Big(\wt\alpha_{i+1},\bar\alpha
\Big) \Theta\Big(\wt\alpha_{i+1},\wt\alpha_i)\Big) 
- \Theta\Big(\bar\alpha, \wt\alpha_{i+1}\Big) 
\Theta\Big(\wt\alpha_i ,\wt\alpha_{i+1}\Big)\bigg] \nn
&& + 
\sum_{k=1}^{R-P-1} \sum_{I_1=P+1}^{R-1}
\sum_{I_2=I_1+1}^{R-1} \sum_{I_3=I_2+1}^{R-1} \cdots
\sum_{I_{k}=I_{k-1}+1}^{R-1} \Bigg\{ \Theta(\bar\alpha, \alpha_{I_1})
\prod_{\ell=1}^{k-1} 
\Theta\Big(\wt\alpha_{I_\ell} , \wt\alpha_{I_{\ell+1}}\Big)\nn
&&\qquad \prod_{i=P}^{I_{1}-2}
\bigg[ \Theta\Big(\wt\alpha_{i+1},\bar\alpha
\Big) \Theta\Big(\wt\alpha_{i+1},\wt\alpha_i\Big) 
- \Theta\Big(\bar\alpha, \wt\alpha_{i+1}\Big) 
\Theta\Big(\wt\alpha_i ,\wt\alpha_{i+1}\Big)\bigg]
\nn
&& \qquad \prod_{\ell=1}^{k-1} 
\prod_{i=I_\ell}^{I_{\ell+1}-2}
\bigg[ \Theta\Big(\wt\alpha_{i+1},\wt\alpha_{I_\ell}
\Big) \Theta\Big(\wt\alpha_{i+1},\wt\alpha_i)\Big) 
- \Theta\Big(\wt\alpha_{I_\ell}, \wt\alpha_{i+1}\Big) 
\Theta\Big(\wt\alpha_i ,\wt\alpha_{i+1}\Big)\bigg]\nn
&& \qquad \prod_{i=I_{k}}^{R-2}
\bigg[ \Theta\Big(\wt\alpha_{i+1},\wt\alpha_{I_k}
\Big) \Theta\Big(\wt\alpha_{i+1},\wt\alpha_i)\Big) 
- \Theta\Big(\wt\alpha_{I_k}, \wt\alpha_{i+1}\Big) 
\Theta\Big(\wt\alpha_i ,\wt\alpha_{i+1}\Big)\bigg]
\Bigg\}\Bigg\}\, .
\een
In this expression $k=p-1$ denotes the total number of $B_a$'s other than $B_1$.
The first term represents the $k=0$ term where there is
a single set $B_1$ containing all the elements from
1 to $R-1$, and the product of the $\Theta$'s account for the
constraint \refb{emag5}. In the other terms
the $I_\ell$'s mark the beginning of the set $B_{\ell+1}$ as in \refb{edefbpart}.
The product of the $\Theta$'s in the second line of \refb{enetc} impose the
constraints \refb{emag4} and the 
$\Theta$'s in the last three lines of \refb{enetc}
impose the constraints \refb{emag5}. The $-$ sign between the two
terms in the first line and the last three lines
originate from the
$(-1)^{\sum_{a=1}^p \sum_{i=I_{a-1}}^{I_a-1}\Theta(\sigma(i)-\sigma(i+1))}$ 
factor in 
\refb{e519}. It takes care of the contribution from the pairs $(i,i+1)$
for $P\le i\le R-1$, but not for values of $i$ outside this range ({\it e.g.} for 
$1\le i<P$) which has been included as an overall factor at the beginning
of \refb{enetc}.

Using the fact that $\Theta(\wt\alpha_i, \wt\alpha_j)
=\Theta(\sigma(i) -\sigma(j))$ where the second $\Theta$ denotes
an ordinary step function, we can convert
\refb{enetc} into a purely combinatoric expression as follows.
Let $k_0$ be the integer
for which $\alpha_{k_0} < \bar\alpha < \alpha_{k_0+1}$. We now
define $c$ to be any number between $k_0$ and $k_0+1$. 
Then $\Theta(\wt\alpha_i, \bar\alpha) =
\Theta(\sigma(i)-c)$ and we may rewrite \refb{enetc} as
\ben \label{enetcnew}
&& (-1)^{\sum_{i=1}^{P-1} \Theta(\sigma(i)-\sigma(i+1))}\, \nn &&
\Bigg[ \prod_{i=P}^{R-2} \bigg[ \Theta\Big(\sigma(i+1)- c
\Big) \Theta\Big(\sigma(i+1)-\sigma(i)\Big) 
- \Theta\Big(c-\sigma(i+1)\Big) \Theta
\Big(\sigma(i) -\sigma(i+1)\Big)\bigg] \nn
&& +
\sum_{k=1}^{R-P-1} \sum_{I_1=P+1}^{R-1}
\sum_{I_2=I_1+1}^{R-1} \sum_{I_3=I_2+1}^{R-1} \cdots
\sum_{I_{k}=I_{k-1}+1}^{R-1} \Bigg\{\Theta(c-\sigma(I_1)) \prod_{\ell=1}^{k-1} 
\Theta\Big(\sigma(I_\ell) - \sigma(I_{\ell+1})\Big)\nn
&&
\prod_{i=P}^{I_{1}-2}
\bigg[ \Theta\Big(\sigma(i+1)
-c\Big) \Theta\Big(\sigma(i+1)-\sigma(i)\Big) 
- \Theta\Big(c - \sigma(i+1)\Big) \Theta\Big(\sigma(i) -\sigma(i+1)\Big)\bigg]
\nn
&&\prod_{\ell=1}^{k-1} 
\prod_{i=I_\ell}^{I_{\ell+1}-2}
\bigg[ \Theta\Big(\sigma(i+1)- \sigma(I_\ell)
\Big) \Theta\Big(\sigma(i+1)-\sigma(i)\Big) 
- \Theta\Big(\sigma(I_\ell)-\sigma(i+1)\Big) \Theta
\Big(\sigma(i) -\sigma(i+1)\Big)\bigg]\nn
&& \prod_{i=I_k}^{R-2} \bigg[ \Theta\Big(\sigma(i+1)- \sigma(I_k)
\Big) \Theta\Big(\sigma(i+1)-\sigma(i)\Big) 
- \Theta\Big(\sigma(I_k)-\sigma(i+1)\Big) \Theta
\Big(\sigma(i) -\sigma(i+1)\Big)\bigg]
\Bigg\}\Bigg]\, .\nn
\een
We now make use of the identity --proved in appendix \ref{sb} -- that for any
function $f(i)$ satisfying $f(i)\ne f(j)$ for $i\ne j$ and any constant $c$, we have
\ben \label{eid1}
&& \prod_{i=1}^{N-1} \bigg[ \Theta\Big( f (i+1)- c
\Big) \Theta\Big( f (i+1)- f (i)\Big) 
- \Theta\Big(c- f (i+1)\Big) \Theta
\Big( f (i) - f (i+1)\Big)\bigg]\nn
&& +\sum_{k=1}^{N-1} \sum_{I_1=2}^{N}
\sum_{I_2=I_1+1}^{N} \sum_{I_3=I_2+1}^{N} \cdots
\sum_{I_{k}=I_{k-1}+1}^{N} \Bigg\{\Theta(c- f (I_1)) \prod_{\ell=1}^{k-1} 
\Theta\Big( f (I_\ell) -  f (I_{\ell+1})\Big)\nn
&&
\prod_{i=1}^{I_{1}-2}
\bigg[ \Theta\Big( f (i+1)
-c\Big) \Theta\Big( f (i+1)- f (i)\Big) 
- \Theta\Big(c -  f (i+1)\Big) \Theta\Big( f (i) - f (i+1)\Big)\bigg]
\nn
&&\prod_{\ell=1}^{k-1} 
\prod_{i=I_\ell}^{I_{\ell+1}-2}
\bigg[ \Theta\Big( f (i+1)-  f (I_\ell)
\Big) \Theta\Big( f (i+1)- f (i)\Big) 
- \Theta\Big( f (I_\ell)- f (i+1)\Big) \Theta
\Big( f (i) - f (i+1)\Big)\bigg]\nn
&& \prod_{i=I_k}^{N-1} \bigg[ \Theta\Big( f (i+1)-  f (I_k)
\Big) \Theta\Big( f (i+1)- f (i)\Big) 
- \Theta\Big( f (I_k)- f (i+1)\Big) \Theta
\Big( f (i) - f (i+1)\Big)\bigg]
\Bigg\}\nn
&=& \prod_{i=1}^{N-1} \Theta( f (i+1)- f (i))\, .
\een
It is easy to see that except for the factor in the first line of 
\refb{enetcnew}, the left hand side
\refb{eid1} reduces to
\refb{enetcnew} under the identification
$N=R-P$, $f(i)=\sigma(i+P-1)$ and a renaming of the variables
$I_\ell$ to $I_\ell - P+1$. Thus \refb{enetcnew} becomes
\be \label{ered1}
 (-1)^{\sum_{i=1}^{P-1} \Theta(\sigma(i)-\sigma(i+1))}\,
 \prod_{i=P}^{R-2} \Theta(\sigma(i+1) -\sigma(i))\, .
\ee
The product of the $\Theta$'s
coincide with the first set of conditions given in \refb{eallowedfin}.

The case where all the $\wt\alpha_i$'s
for $1\le i\le R-1$ are almost zero requires special treatment.
In this case \refb{emag} holds all the way to $\wt\alpha_1$ with
$|\wt\alpha_1|$ being the largest and \refb{emag2} holds for
$1\le \ell\le \ell'\le R-1$. Thus we have $\delta^{(1)}\simeq \wt\alpha_1$
and the analog of \refb{enetcnew} takes the form
\ben \label{enetcP}
&&\prod_{i=1}^{R-2}
\bigg[ \Theta\Big(\sigma(i+1)
-\sigma(1)\Big) \Theta\Big(\sigma(i+1)-\sigma(i)\Big) 
- \Theta\Big(c - \sigma(i+1)\Big) \Theta\Big(\sigma(i) -\sigma(i+1)\Big)\bigg]
\nn
&& + \sum_{k=1}^{R-2} \sum_{I_1=2}^{R-1}
\sum_{I_2=I_1+1}^{R-1} \sum_{I_3=I_2+1}^{R-1} \cdots
\sum_{I_{k}=I_{k-1}+1}^{R-1} \Bigg\{\Theta(\sigma(1)-\sigma(I_1)) 
\prod_{\ell=1}^{k-1} 
\Theta\Big(\sigma(I_\ell) - \sigma(I_{\ell+1})\Big)\nn
&&
\prod_{i=1}^{I_{1}-2}
\bigg[ \Theta\Big(\sigma(i+1)
-\sigma(1)\Big) \Theta\Big(\sigma(i+1)-\sigma(i)\Big) 
- \Theta\Big(\sigma(1) - \sigma(i+1)\Big) \Theta\Big(\sigma(i) -\sigma(i+1)\Big)\bigg]
\nn
&&\prod_{\ell=1}^{k-1} 
\prod_{i=I_\ell}^{I_{\ell+1}-2}
\bigg[ \Theta\Big(\sigma(i+1)- \sigma(I_\ell)
\Big) \Theta\Big(\sigma(i+1)-\sigma(i)\Big) 
- \Theta\Big(\sigma(I_\ell)-\sigma(i+1)\Big) \Theta
\Big(\sigma(i) -\sigma(i+1)\Big)\bigg]\nn
&& \prod_{i=I_k}^{R-2} \bigg[ \Theta\Big(\sigma(i+1)- \sigma(I_\ell)
\Big) \Theta\Big(\sigma(i+1)-\sigma(i)\Big) 
- \Theta\Big(\sigma(I_\ell)-\sigma(i+1)\Big) \Theta
\Big(\sigma(i) -\sigma(i+1)\Big)\bigg]
\Bigg\}\,. \nn
\een
This is identical to the left hand side of \refb{eid1} with $c$ replaced by
$\sigma(1)$, $f(i)$ replaced by $\sigma(i)$ and $N$ replaced by
$R-1$. Thus the result is
\be \label{ered2}
\prod_{i=1}^{R-2} \Theta(\sigma(i+1) -\sigma(i))\, .
\ee

We can analyze the contribution from the right hand side of $R$ by
summing over all possible choices of $q$ and
$J_1, \cdots J_{q-1}$. 
Since this is related to the previous analysis by a reversal of permutation
the result can be read out from our previous analysis.
However there are two important effects which
need to be taken into account. First of all
\refb{e519} has a factor of $(-1)^q$ which in the present context
will turn into $(-1)^{k+1}$. Second in order to convert this problem to the
previous case we need to flip the sign of each term inside $[~]$ in \refb{enetc}
since the pair $(i,i+1)$ goes over to $(i'+1,i')$ for some $i'$ under
permutation reversal. This gives a factor of $(-1)^f$ where 
$f$ essentially counts the total number of nearest neighbor
pairs between 
$R+1$ and $Q$ except the links which connect
the end point of $C_a$ to
the starting point of $C_{a-1}$ for $2\le a\le q$. 
Thus we have $f=Q-R-1-(q-1)= Q-R-k-1$
and hence the net factor is 
\be \label{efactorof}
(-1)^{f+k+1}= (-1)^{Q-R}\, .
\ee
The result of summing over the locations of
$J_1,\cdots J_{q-1}$ in the range $R$ to $Q$ is now  given by
\be \label{ered3}
(-1)^{\sum_{i=Q}^n \Theta(\sigma(i)-\sigma(i+1))} 
(-1)^{Q-R}\prod_{j=R+1}^{Q-1} \Theta(\sigma(j) - \sigma(j+1))\, ,
\ee 
both when $Q<n+1$ and when $Q=n+1$.
This 
shows that between $R$ and $Q$ the $\sigma(j)$'s must 
form a decreasing
sequence. The first factor in \refb{ered3} is the contribution
from the points between $Q$ and $n+1$.

Eqs.\refb{ered1} and \refb{ered3} and the fact that
$\sigma(R)=n+1>\sigma(R-1),\sigma(R+1)$
together give
\be \label{eallowedfinks}
\hbox{$\sigma(k+1)>\sigma(k)$ for $P\le k\le R-1$}, \quad
\hbox{$\sigma(k+1)<\sigma(k)$ for $R\le k\le Q-1$}\, .
\ee
which precisely correspond to
the condition \refb{eallowedfin} for $g_{\rm higgs}$ to
be non-vanishing.
The net contribution to ${\rm N}^{(n+1)}_{\rm KS}$ for a
configuration satisfying \refb{eallowedfinks} and the other
conditions described in \S\ref{econper} is now given by the product of
\refb{ered1} and \refb{ered3} (with $P$ replaced by $1$ or $Q$ replaced
by $n+1$ in special cases).
The result is
\be \label{enksign}
{\rm N}^{(n+1)}_{\rm KS}(\{\alpha_i\};\sigma) = (-1)^{r(\sigma)}, \quad
r(\sigma)=Q-R+ \sum_{i=1}^{P-1}  \Theta(\sigma(i)-\sigma(i+1))
+ \sum_{i=Q}^{n}  \Theta(\sigma(i)-\sigma(i+1)) \, .
\ee
This is in perfect agreement with \refb{enmpspred}.

Finally we note that the very special cases when $R$ itself lies at either
end can also be easily incorporated in this analysis. For example
if $R=1$ we simply need to drop the $\sum_{i=1}^{P-1}  
\Theta(\sigma(i)-\sigma(i+1))$ term and if $R=n+1$ we shall need to
drop the $Q-R+ \sum_{i=Q}^{n}  \Theta(\sigma(i)-\sigma(i+1))$ term.
These are also in agreement with the corresponding formula
\refb{enmpspred} for ${\rm N}^{(n+1)}_{\rm higgs}$.

\sectiono{An alternative approach to proving the equivalence of the Higgs branch
and KS wall crossing formul\ae} \label{salter}

In \S\ref{s4} and \S\ref{s5} we gave a proof of the equality of $g_{\rm KS}$ and
$g_{\rm higgs}$ for generic choice of the arguments $\alpha_1,\cdots \alpha_n$.
This proof used the definition of $g_{\rm KS}$ given in \S\ref{s2} based on
the universal enveloping algebra of the Lie algebra \refb{eliealg}. The
referee suggested a simpler approach based on the quantum torus algebra
\be \label{eat1}
e_{\gamma}  e_{\gamma'} = (y-y^{-1})^{-1}\, 
(-y)^{\langle\gamma, \gamma' \rangle}
e_{\gamma+\gamma'}\, ,
\ee
which provides a representation of \refb{eliealg}. The KS wall crossing
formula takes the form:
\be \label{aet2}
\prod_{\gamma \, \, \hbox{\small clockwise}} 
\exp\left[\bar\Omega^-_{\rm ref}(\gamma,y)\, e_\gamma\right]
= \prod_{\gamma \,\,  \hbox{\small anti-clockwise}} 
\exp\left[\bar\Omega^+_{\rm ref}(\gamma,y)\, e_\gamma\right]\, ,
\ee
where $\prod_{\gamma \, \, \hbox{\small clockwise}}$ 
($\prod_{\gamma \, \, \hbox{\small anti-clockwise}}$) denotes product over
all vectors $\gamma\in\Lambda$, 
and the terms in the product arranged such that as we move from the
left to the right the corresponding $\gamma$'s are arranged clockwise 
(anti-clockwise) in the two
dimensional plane. Expanding both sides using \refb{eat1}, and collecting the
coefficient of $e_\gamma$ from each side, we get
\ben \label{eat3}
\sum_{m=1}^\infty \sum_{\beta_1, \cdots \beta_m\in\Lambda\atop
\beta_1+\cdots \beta_m=\gamma, \, \beta_1\le \beta_2\le \cdots \le \beta_m}
{1\over |{\rm Aut} (\{\beta_1,\cdots \beta_m\})|}
(y-y^{-1})^{-m} (-y)^{\sum_{i<j} \beta_{ij}}
\prod_{i=1}^m \bar\Omega^-_{\rm ref}(\beta_i,y) \nonumber \\
= \sum_{n=1}^\infty  \sum_{\alpha_1, \cdots \alpha_n\in\Lambda\atop
\alpha_1+\cdots \alpha_n=\gamma, \, \alpha_1\le \alpha_2\le \cdots \le \alpha_n}
{1\over |{\rm Aut} (\{\alpha_1,\cdots \alpha_n\})|}(y-y^{-1})^{-n} (-y)^{-\sum_{i<j} \alpha_{ij}}
\prod_{i=1}^n \bar\Omega^+_{\rm ref}(\alpha_i,y)\, ,
\een
where by $\beta_i=\beta_j$ we mean $\beta_i$ is either equal or parallel
to $\beta_j$.
We now substitute in the left hand side of this equation
the expression for $\bar\Omega^-_{\rm ref}(\beta_i,y)$ in terms 
of $\bar\Omega^+_{\rm ref}(\alpha_i,y)$'s using eqs.\refb{efirst3} with
$g_{\rm ref}$ replaced by $g_{\rm KS}$. 
This gives
\ben \label{eat3.5}
\sum_{m=1}^\infty \sum_{\beta_1, \cdots \beta_m\in\Lambda\atop
\beta_1+\cdots \beta_m=\gamma, \, \beta_1\le \beta_2\le \cdots \le \beta_m}
{1\over |{\rm Aut} (\{\beta_1,\cdots \beta_m\})|}
(y-y^{-1})^{-m} (-y)^{\sum_{i<j} \beta_{ij}} \qquad \qquad  \qquad
 \quad \nonumber \\
\prod_{k=1}^m 
\sum_{s_k\geq 1 }\, 
  \sum_{\hbox{\small \tiny unordered set}\, \alpha^{(k)}_1,\dots, \alpha^{(k)}_{s_k}
  \in\Lambda \atop
\alpha^{(k)}_1+\dots +\alpha^{(k)}_{s_k}=\beta_k}\, 
\frac{g_{\rm KS}(\alpha^{(k)}_1,\dots, \alpha^{(k)}_{s_k})}{|{\rm Aut}
(\{\alpha^{(k)}_1,\dots, \alpha^{(k)}_{s_k}\})|}\, 
 \prod\nolimits_{i=1}^{s_k} \bOm_{\rm ref}^+(\alpha^{(k)}_i,y)  \qquad  \qquad
\nonumber \\
= \sum_{n=1}^\infty  \sum_{\alpha_1, \cdots \alpha_n\in\Lambda\atop
\alpha_1+\cdots \alpha_n=\gamma, \, \alpha_1\le \alpha_2\le \cdots \le \alpha_n}
{1\over |{\rm Aut} (\{\alpha_1,\cdots \alpha_n\})|}(y-y^{-1})^{-n} (-y)^{-\sum_{i<j} \alpha_{ij}}
\prod_{i=1}^n \bar\Omega^+_{\rm ref}(\alpha_i,y)\, .
\een
Comparing the coefficient of $\prod_{i=1}^n \bar\Omega^+_{\rm ref}(\alpha_i,y)$
in two sides of this equation we can get a set of recursion relations involving 
$g_{\rm KS}(\alpha_1, \cdots \alpha_n)$ from which we can
determine $g_{\rm KS}(\alpha_1, \cdots \alpha_n)$. Thus
proving the equivalence of $g_{\rm KS}$ and $g_{\rm higgs}$ is equivalent to
checking if $g_{\rm higgs}$ satisfies the same set of relations \refb{eat3.5}
as $g_{\rm KS}$. For this we replace $g_{\rm KS}$ in \refb{eat3.5} by
the expression for $g_{\rm higgs}$
given in \refb{higgsfnewest} for generic arguments
and try to verify the resulting equation. 
Comparing the coefficients of 
$\prod_{i=1}^n \bar\Omega^+_{\rm ref}(\alpha_i,y)$ on both sides of this equation
for generic $\alpha_i$'s for which 
\be \label{ealporder}
\alpha_1<\alpha_2<\cdots <\alpha_n
\ee
and
$\sum_{i\in A}\alpha_i$ and $\sum_{i\in B}\alpha_i$
are different from each other for any choice of non-overlapping sets $A$
and $B$ of $\{1,2,\cdots n\}$, we get:
\ben \label{eat4}
&& \sum_{\hbox{\tiny permutations} \, \sigma\atop
\hbox{\tiny of}\, \, 1,2,\cdots n} \sum_{m=1}^n \sum_{n_1,\cdots n_m\atop 0\equiv n_0 
< n_1 < n_2<\cdots < n_m\equiv n}
(y-y^{-1})^{-m} \, \left\{\prod_{k=1}^{m-1} \Theta\left(\beta_{k+1}, \beta_k
\right) \right\}
(-y)^{\sum_{k=1}^{m-1} \sum_{l=k+1}^m \langle \beta_k, \beta_l\rangle
}\nonumber \\ && 
(-1)^{n-m} (y - y^{-1})^{m-n} (-y)^{\sum_{k=1}^m \sum_{i=n_{k-1}+1}^{n_k-1} 
\sum_{j=i+1}^{n_k} \alpha_{\sigma(i)\sigma(j)}} (-1)^{\sum_{k=1}^m 
\sum_{i=n_{k-1}+1}^{n_k-1} \Theta\left(\alpha_{\sigma(i)}, \alpha_{\sigma(i+1)}\right)}
\nonumber \\ &&
\prod_{k=1}^m \left\{ \prod_{i=n_{k-1}+1\atop \sigma(i)>\sigma(i+1)}^{n_k-1} 
\Theta\left(\sum_{j=n_{k-1}+1}^i \alpha_{\sigma(j)}, 
\beta_k\right) 
\prod_{i=n_{k-1}+1\atop \sigma(i)<\sigma(i+1)}^{n_k-1}  \Theta\left(
\beta_k, \sum_{j=n_{k-1}+1}^i \alpha_{\sigma(j)}
\right)\right\} \nonumber \\
&&= (y-y^{-1})^{-n} \, (-y)^{-\sum_{i<j} \alpha_{ij}} \, , \nonumber \\
&& \beta_k \equiv \sum_{j=n_{k-1}+1}^{n_{k}} \alpha_{\sigma(j)}\, .
\een
The sum over $\sigma$ runs over all permutations of $\{1,2,\cdots n\}$.
For a fixed $\sigma$, the different choices of the integers $n_k$ correspond
to different partitioning of the ordered set $\{\alpha_{\sigma(1)}, \cdots
\alpha_{\sigma(n)}\}$. The sum of the vectors inside the partitions from
left to right are given by  $\beta_1,\cdots \beta_m$, satisfying the
constraint $\beta_1<\beta_2<\cdots <\beta_m$ as in the left hand side of
\refb{eat3.5}.\footnote{Note that since we are taking the $\alpha_i$'s to be
generic, we do not consider the case where some of the $\beta_k$'s are
equal or parallel to each other.}
After cancelling the $(y-y^{-1})^{-n}$ factors from the two sides,
we note that the net power of $(-y)$ on the left hand side is given by
$\sum_{i=1}^{n-1}\sum_{j=i+1}^n \alpha_{\sigma(i)\sigma(j)}$, \i.e.\ the power
of $(-y)$ is determined only by the permutation $\sigma$ and is independent
of the choice of the integers $m$ and
$n_1,\cdots n_m$ which  partitions the vectors into $\{\beta_1,\cdots \beta_m\}$.
Comparing the different powers of $y$ on the two sides of
\refb{eat4} we now get
\ben \label{eat5}
&&\sum_{m=1}^n \sum_{n_1,\cdots n_m\atop 0\equiv n_0 
< n_1 < n_2<\cdots < n_m\equiv n} 
\left\{\prod_{k=1}^{m-1}  \Theta\left(\beta_{k+1}, \beta_k
\right)\right\}
(-1)^{n-m+\sum_{k=1}^m 
\sum_{i=n_{k-1}+1}^{n_k-1} \Theta\left(\alpha_{\sigma(i)}, \alpha_{\sigma(i+1)}\right)}
\nonumber \\ &&
\prod_{k=1}^m \left\{ \prod_{i=n_{k-1}+1\atop \sigma(i)>\sigma(i+1)}^{n_k-1} 
\Theta\left(\sum_{j=n_{k-1}+1}^i \alpha_{\sigma(j)}, 
\beta_k\right) 
\prod_{i=n_{k-1}+1\atop \sigma(i)<\sigma(i+1)}^{n_k-1}  \Theta\left(
\beta_k, \sum_{j=n_{k-1}+1}^i \alpha_{\sigma(j)}
\right)\right\} \nonumber \\
&&= \cases{
1 \quad \hbox{for $\sigma(1,2,\cdots n)=(n,n-1,\cdots, 1)$}\cr
0 \quad \hbox{otherwise}
}\een
For a fixed $\sigma$, we shall refer to the choice of $m$ and $\{n_1,\cdots n_m\}$
for which the summand is non-vanishing as an allowed partition.

Our goal now is to prove \refb{eat5}. We begin with the case
$\sigma(1,2,\cdots n)=(n,n-1,\cdots 1)$. In this case the only way to avoid a
vanishing
contribution from the $\prod_{k=1}^{m-1}  \Theta\left(\beta_{k+1}, \beta_k
\right)$ factor is to choose $m=1$, $n_1=n$.
In this case
the
condition \refb{ealporder}
tells us that
\be \label{eat10}
\alpha_{\sigma(i)}> \alpha_{\sigma(i+1)}, \quad 
\sum_{j=1}^i \alpha_{\sigma(j)} > \beta_1=\alpha_1+\cdots \alpha_n\, .
\ee
Hence the
product of the step functions given in the last line on the left hand side 
of \refb{eat5} is
1 since for all $i$ we have $\sigma(i)> \sigma(i+1)$ and
$\sum_{j=1}^i \alpha_{\sigma(j)} > \beta_1$.
The sign of the term is
$(-1)^{n-1+n-1}=1$. Thus the result is 1 in agreement with the right hand
side of \refb{eat5}.

To deal with the case of other permutations, we note first of all that the right hand
side of \refb{eat5} is independent of the $\alpha_i$'s. 
Thus we need to show that the left hand side 
of this equation must also be invariant under deformations of the $\alpha_i$'s
 as long as
we preserve the order \refb{ealporder}. This is not manifest, 
{\it e.g.} during such deformations of the $\alpha_i$'s, in a given term in
\refb{eat5}
$\beta_k$ and $\beta_{k+1}$ defined in \refb{eat4} may go from $\beta_k<\beta_{k+1}$ to 
$\beta_k>\beta_{k+1}$ and as a result $\Theta(\beta_{k+1},\beta_k)$ may jump from 1 to 0.
Thus if we choose 
$n_1,\cdots n_m$ such that initially  we have $\beta_k<\beta_{k+1}$, and
the restrictions imposed by the
various other step functions in \refb{eat5} are satisfied so that we have a
non-zero contribution, during the deformation 
we may arrive at $\beta_k>\beta_{k+1}$ so that this term ceases to
contribute.
Thus to show that the left
hand side of \refb{eat5} is unchanged during such a deformation we must
identify another contribution that either ceases to exist or begins to exist 
when $\beta_k$ and $\beta_{k+1}$ switches order,
compensating for the change caused due to the previous effect. We consider two
cases separately: $\alpha_{\sigma(n_k)} < \alpha_{\sigma(n_k+1)}$ and 
$\alpha_{\sigma(n_k)} > \alpha_{\sigma(n_k+1)}$.
In the first case it is easy to see that as long as $\beta_k<\beta_{k+1}$, the configuration
with $m\to m-1$, and the $\beta_k$'s chosen as $\{\beta_1, \cdots \beta_{k-1},
\beta_{k}+\beta_{k+1}, \beta_{k+2}, \cdots \beta_m\}$ also contributes to
the left hand side of \refb{eat4}, and furthermore, this configuration also ceases 
to contribute when we cross over to the side where $\beta_k>\beta_{k+1}$.
Furthermore for $\beta_k<\beta_{k+1}$ the contribution from these two configurations
 have opposite signs so that the sum of the two terms vanishes
 and there is no net change in the left hand side of
\refb{eat5} as we pass from $\beta_k<\beta_{k+1}$ to $\beta_k>\beta_{k+1}$. 
On the other hand if $\alpha_{\sigma(n_k)} > \alpha_{\sigma(n_k+1)}$, then the
second configuration does not contribute for $\beta_k<\beta_{k+1}$ but does
contribute when $\beta_k>\beta_{k+1}$ and the sign of the second
contribution for $\beta_k>\beta_{k+1}$
is the same as that of the first configuration for $\beta_k<\beta_{k+1}$.
Thus again the net contribution remains unchanged during this process.

The other possible source of jump in the left hand  side of
\refb{eat5} is when $\sum_{j=n_{k-1}+1}^i \alpha_{\sigma(j)}$
crosses $\beta_k$ for some $(i,k)$ during the deformation. In this case 
we analyze the contribution from the original partition together with that of 
another partition corresponding to $m\to m+1$, with the $\beta_i$'s given by
$$\{\beta_1, \cdots \beta_{k-1}, \sum_{j=n_{k-1}+1}^i \alpha_{\sigma(j)}, 
\sum_{j=i+1}^{n_k}\alpha_{\sigma(j)}, \beta_{k+1}, \cdots \beta_{m}\}.$$
The analysis of the net jump of the left hand side of \refb{eat5} from these two
terms  is
the same as the one given in the last paragraph, with the roles of the first and
the second terms getting exchanged.

This shows that the left hand side of \refb{eat5} is unchanged under continuous deformation
of the $\alpha_i$'s preserving \refb{ealporder}. 
Armed with this result we shall now try to prove \refb{eat5} using 
the method of induction, \i.e.\ we shall assume that the
result is valid for $(n-1)$ $\alpha_i$'s and then try to prove it for $n$ $\alpha_i$'s. 
Using the invariance of the left hand side of \refb{eat5} under deformations of the
$\alpha_i$'s preserving \refb{ealporder}, 
we shall choose $\alpha_{\sigma(n)}$ to be
small in magnitude compared to all the other $\alpha_{\sigma(i)}$'s. 
In this case
all the step functions in \refb{eat5}, except for the ones which contain
$\alpha_{\sigma(n)}$ as one of its arguments, reduce to those for 
the case of $(n-1)$ vectors $\alpha_{\sigma(1)}, \cdots \alpha_{\sigma(n-1)}$.
Thus the possible allowed
partitions are of the form:
\be \label{eat6}
\{\beta_1, \cdots \beta_m+\alpha_{\sigma(n)}\}, \quad
\{\beta_1, \cdots \beta_m,\alpha_{\sigma(n)}\}\, ,
\ee
where $\{\beta_1,\cdots \beta_m\}$ is an allowed partition for
$(n-1)$ vectors $\alpha_{\sigma(1)},\cdots \alpha_{\sigma(n-1)}$. 
We now consider two possibilities:
\begin{enumerate}
\item $\alpha_{\sigma(n)}> \alpha_{\sigma(n-1)}$: In this case 
by examining the additional step functions
which involve $\alpha_{\sigma(n)}$ as one of the arguments we see that
both the partitions given in \refb{eat6}
can contribute only when $\alpha_{\sigma(n)}>\beta_m$,
and they contribute with opposite sign. Thus we conclude that the net
contribution vanishes.
\item $\alpha_{\sigma(n)}< \alpha_{\sigma(n-1)}$: In this case the first partition
contributes when $\alpha_{\sigma(n)}<\beta_m$ and the second partition
contributes when $\alpha_{\sigma(n)}>\beta_m$. Both contributions are equal,
and so we get a non-vanishing contribution that is independent of whether
$\alpha_{\sigma(n)}<\beta_m$ of $\alpha_{\sigma(n)}>\beta_m$. This allows
us to sum over all partitions $\{\beta_1,\cdots \beta_m\}$ of $\alpha_{\sigma(1)},
\cdots \alpha_{\sigma(n-1)}$ freely, allowing us to use \refb{eat5} for $(n-1)$
vectors. This leaves us with the result that the only
permutation for which we have a non-vanishing contribution is 
$\alpha_{\sigma(1)}>\alpha_{\sigma(2)}>\cdots >\alpha_{\sigma(n-1)}$.
Combining this with the result that $\alpha_{\sigma(n)}<\alpha_{\sigma(n-1)}$
we see that the only permutation for which the result is non-zero is
$\sigma(1,2,\cdots n)=(n,n-1,\cdots 1)$. This is the desired result.
\end{enumerate}

\sectiono{Equivalence of `higgs' and `coulomb' branch wall crossing formul\ae} \label{coulomb}

Ref.\cite{Manschot:2010qz} also proposed a different prescription for computing
$g_{\rm ref}(\alpha_1,\cdots,\alpha_n; y)$ 
called the `coulomb branch formula'. The formula is
similar to \refb{higgsfnewest}, but with ${\rm N}^{(n)}_{\rm higgs}(\{\alpha_i\};\sigma)$ 
replaced by an apparently
different quantity which we shall denote by 
${\rm N}^{(n)}_{\rm coulomb}(\{\alpha_i\};\sigma)$. 
The prescription for computing ${\rm N}^{(n)}_{\rm coulomb}(\{\alpha_i\};\sigma)$
associated with a given permutation $\sigma$ is as follows:
\begin{enumerate}
\item Let us define $\wt\alpha_i\equiv\alpha_{\sigma(i)}$ 
as usual. Now for a given
permutation $\sigma$ we consider a function $W$ of $n$ real variables
$ x_1,\cdots  x_n$ ordered as $ x_1< x_2<\cdots <  x_n$:
\be \label{ecou1}
W=-\sum_{i<j} \langle \wt\alpha_i, \wt\alpha_j\rangle \log| x_i- x_j|
+\Lambda \sum_{i,j=1}^n \langle \wt\alpha_i, \wt\alpha_j\rangle  x_j\, ,
\ee
where $\Lambda$ is a positive constant. It will be useful to think of $W$ as
the potential for $n$ one dimensional particles positioned at
$ x_1,\cdots  x_n$. $W$ is invariant under simultaneous translation of all
$ x_i$'s by a constant.
\item ${\rm N}^{(n)}_{\rm coulomb}(\{\alpha_i\};\sigma)$ is non vanishing 
only if there is 
an extremum of $W$ with respect to the variables $ x_2, \cdots  x_n$ at fixed 
$ x_1=0$:
\be \label{ewextreme}
{\p W\over \p x^i}=0 \quad \hbox{for $2\le i\le n$}\, .
\ee
This corresponds to an equilibrium configuration of the $n$ particles
and fixing $ x_1=0$ (or any other fixed value) is possible due to translation
invariance of $W$ mentioned above.
When this condition is satisfied ${\rm N}^{(n)}_{\rm coulomb}(\{\alpha_i\};\sigma)$ takes value
1 or $-1$, with the sign given by
the sign of $\det M$ at the extremum, where $M_{ij}=
(\p^2 W/\p  x_i \p  x_j)$, $i,j=2, \cdots n$. If there is more
than one extremum then we have to sum over these extrema with
weight factors ${\rm sign}(\det M)$ associated with these different extrema.
\end{enumerate}

Our goal in this section will be
to prove the equality of ${\rm N}^{(n)}_{\rm higgs}(\{\alpha_i\};\sigma)$ and
${\rm N}^{(n)}_{\rm coulomb}(\{\alpha_i\};\sigma)$. We first note that the prescription for computing
${\rm N}^{(n)}_{\rm coulomb}$ given above is invariant under small deformations
of $\{\alpha_i\}$ under which the extrema change their positions or (dis)appear
in pairs by merger or pair creation, but does not (dis)appear singly. 
The latter may occur if during the deformation a nearest neighbor pair
$( x_i,  x_{i+1})$ at the extremum
approach each other so that beyond the point
of merger of $ x_i$ and $ x_{i+1}$ the extremum ceases to exist, or if at the
extremum
a set of $ x_i$'s get separated from the rest by an infinite distance beyond which
the extremum ceases to exist. Now since as $ x_i\to  x_{i+1}$ the dominant term
in $\p W/\p  x_i$ is given by
\be \label{edom}
\langle\wt\alpha_i, \wt\alpha_{i+1}\rangle \, ( x_{i+1}-  x_i)^{-1}
\ee
we see that this term cannot be cancelled by the contribution from any other term
unless $\langle\wt\alpha_i, \wt\alpha_{i+1}\rangle$ approaches zero. Thus as
long as the deformations preserve the relative ordering of the $\alpha_i$'s
so that for no pair $\langle\alpha_i, \alpha_j\rangle$ passes through 0, we avoid
the first possibility. 
To examine the second possibility let us suppose that at the extremum
the subset of points $\{ x_i, k+1\le i\le n\}$ gets separated from the rest
$\{ x_i, 1\le i\le k\}$ by an infinite distance. 
Since at the extramum $\p W/\p  x_j$ should vanish for
$j\ge 2$, we must have
$\sum_{j=k+1}^n \p W/ \p  x_j=0$. In the large separation limit the second
term in \refb{ecou1} dominates and we get
\be \label{ecou2}
\sum_{j=k+1}^n \p W/ \p  x_j\simeq
\Lambda \sum_{i=1}^n \sum_{j=k+1}^n \langle \wt\alpha_i,\wt\alpha_j\rangle
= \Lambda \langle \bar\alpha, \bar\alpha-\beta_k\rangle
= -\Lambda \langle \bar\alpha, \beta_k\rangle\, ,
\ee
where $\beta_k=\sum_{j=1}^k \wt\alpha_j$
and $\bar\alpha=(\alpha_1+\cdots \alpha_n)$  as before. Thus as long as
this is kept away from zero the extremum of $W$ cannot approach a
configuration where $ x_1,\cdots  x_k$ separates by an infinite distance
from the rest of the $ x_i$'s.
Combining these results we come to the conclusion that if we deform the
$\alpha_i$'s without changing the relative orientation between the
$\alpha_i$'s and the relative orientation between any of the $\beta_k$'s
and $\bar\alpha$, ${\rm N}^{(n)}_{\rm coulomb}(\{\alpha_i\};\sigma)$  
will remain unchanged. These
are the same set of deformations under which 
${\rm N}^{(n)}_{\rm higgs}(\{\alpha_i\};\sigma)$ remains
unchanged.

We are now ready to begin proving the equality of ${\rm N}^{(n)}_{\rm higgs}$ and 
${\rm N}^{(n)}_{\rm coulomb}$. 
This has been checked explicitly for low values on $n$ in
\cite{Manschot:2010qz}. 
We shall use the method of induction \i.e.\ assume that
the equality of the ${\rm N}^{(m)}_{\rm higgs}$ and
$ {\rm N}^{(m)}_{\rm coulomb}$ holds for $m\le n-1$ and then prove the
equality of ${\rm N}^{(n)}_{\rm higgs}$ and 
${\rm N}^{(n)}_{\rm coulomb}$. Let us consider a particular $\sigma$ and the
associated $\wt\alpha_i$'s. We now deform the pair of charges 
$(\wt\alpha_{n-1}, \wt\alpha_n)$ according to the prescription of
\S\ref{s5.2}: $\wt\alpha_{n-1}\to (1+\lambda) \wt\alpha_{n-1}$,
$\wt\alpha_n\to (1+\lambda')\wt\alpha_n$ with
$\lambda\alpha_{n-1}+\lambda'\alpha_n\propto \bar\alpha
$ and $\lambda<0$. 
We can increase $|\lambda|$ without changing the relative
orientations between the $\alpha_i$'s and between the $\beta_k$'s and
$\bar\alpha$ till we encounter one of the following situations: 
either $\wt\alpha_n$ becomes
almost zero (which is equivalent to $\beta_{n-1}$ becoming almost
parallel to $\bar\alpha$) or $\wt\alpha_{n-1}$ becomes almost zero.
{\it In either case the charge that becomes almost zero 
does not affect the equilibrium 
arrangement of the rest of the $(n-1)$ centers obtained by
extremizing $W$ with respect to the $(n-2)$ variables. Thus
the arrangement
of the rest of the centers at the extremum must follow the rules of
${\rm N}^{(n-1)}_{\rm coulomb}(\{\alpha_i\};\sigma)$, which by our ansatz are the
same as those of
${\rm N}^{(n-1)}_{\rm higgs}(\{\alpha_i\};\sigma)$.}
We shall now consider the two possibilities separately.

First suppose that $\wt \alpha_n$ becomes almost zero. In this case
the charges $\wt\alpha_1,\cdots \wt \alpha_{n-1}$
for which ${\rm N}^{(n-1)}_{\rm higgs}$ is non-zero can be found by
replacing $n+1$ by $n-1$ in  \refb{eallowed} 
\be \label{eallowednm1}
\beta_{k} \cases{\hbox{$< \bar\alpha$ for 
$\wt\alpha_{k+1}>\wt\alpha_{k}$}\cr
\hbox{$>\bar\alpha$ for $\wt\alpha_{k+1}<\wt\alpha_{k}$}
}\, , \qquad \hbox{for $1\le k\le n-2$}\, .
\ee
By the assumed equality of ${\rm N}^{(n-1)}_{\rm higgs}$ and
${\rm N}^{(n-1)}_{\rm coulomb}$, we have an equilibrium configuration
of $ x_1,\cdots  x_{n-1}$ iff the charges satisfy \refb{eallowednm1}.
Since the addition of an infinitesimal charge
$\wt\alpha_n$ at $ x_n$ will not disturb the equilibrium configuration
of the other charges, we only need to look for an equilibrium 
position of $x_n$, \i.e.\ an $ x_n$ satisfying
$\p W/\p  x_n=0$.
{}From \refb{ecou1} we see that
\ben \label{ewesee}
W &\to& \Lambda \sum_{i=1}^{n-1} 
\langle \wt\alpha_i, \wt\alpha_n\rangle  x_n 
= \Lambda \langle \beta_{n-1}, \bar\alpha -\beta_{n-1}\rangle  x_n 
=\Lambda \langle \beta_{n-1}, \bar\alpha \rangle  x_n
\quad \hbox{as $ x_n\to\infty$}\nn
W &\to& -
\langle \wt\alpha_{n-1}, \wt\alpha_n\rangle \ln| x_{n-1}- x_n|
\quad \hbox{as $ x_n\to  x_{n-1}$}\, .
\een
At both limits the magnitude of $W$ goes to infinity. Thus if these two
limits have the same sign then $\p W/\p  x_n$ must vanish somewhere
in the range $ x_{n-1}< x_n<\infty$ and
we are guaranteed to have an extremum.
This gives
\ben \label{ecou4}
\wt \alpha_{n-1} > \wt\alpha_n &\hbox{if}& \beta_{n-1}>\bar\alpha\nn
\wt \alpha_{n-1} < \wt\alpha_n &\hbox{if}& \beta_{n-1}<\bar\alpha\, .
\een
If this condition is satisfied then we may have more than one solution, but the
number of solutions is always odd and all except one contribution cancels when we
weigh it by the sign of $\det M$. If \refb{ecou4} does not hold then
we could have even number of solutions but their contribution will
cancel pairwise.

\refb{ecou4} precisely extends \refb{eallowednm1} all the way to $k=n-1$.
Thus we see that the condition for ${\rm N}^{(n)}_{\rm coulomb}(\{\alpha_i\};\sigma)$ to be
non-vanishing coincides with the condition for ${\rm N}^{(n)}_{\rm higgs}(\{\alpha_i\};\sigma)$ to
be non-vanishing. We shall now show that the signs of 
${\rm N}^{(n)}_{\rm coulomb}(\{\alpha_i\};\sigma)$ and ${\rm N}^{(n)}_{\rm higgs}(\{\alpha_i\};\sigma)$ also
agree in this case. 
In the limit when the charge of the $n$-th center is small we can
ignore the off diagonal $M_{ni}$ and $M_{in}$ components
of the matrix $M$ and
express $\det M$ as the product of the
determinant of the first $(n-2)\times (n-2)$
block and $\p^2 W/\p  x_n^2$.
The assumed equality of 
${\rm N}^{(n-1)}_{\rm coulomb}(\{\alpha_i\};\sigma)$ and ${\rm N}^{(n-1)}_{\rm higgs}(\{\alpha_i\};\sigma)$
tell us that the sign of the contribution from the first
$(n-2)\times (n-2)$
block to $\det M$ is given by \refb{higgsfnewest}
\be \label{ecou5}
(-1)^{\sum_{k=1}^{n-2} \Theta(\wt\alpha_i, \wt\alpha_{i+1})}\, .
\ee
Now we see from \refb{ewesee}, \refb{ecou4} that
if $\wt\alpha_{n-1}>\wt\alpha_n$ \i.e.\
$\langle\wt\alpha_{n-1},\wt\alpha_n\rangle <0$, then $W\to -\infty$ as
$ x_n\to  x_{n-1}, \infty$ and hence $\p^2 W/\p  x_n^2$ at the 
extremum is negative. On the other hand 
if $\wt\alpha_{n-1}<\wt\alpha_n$ then $\p^2 W/\p  x_n^2$ at the extremum
is positive. Thus ${\rm N}^{(n)}_{\rm coulomb}(\{\alpha_i\};\sigma)$
is obtained by multiplying \refb{ecou5} by a factor of
$(-1)^{\Theta(\wt\alpha_{n-1}, \wt\alpha_{n})}$. This reproduces the
formula for ${\rm N}^{(n)}_{\rm higgs}$ given in
\refb{higgsfnewest}, showing the equality of
${\rm N}^{(n)}_{\rm higgs}$ and ${\rm N}^{(n)}_{\rm coulomb}$.

Next we consider the case when $\wt\alpha_{n-1}$ becomes almost zero
at the end of the deformation. 
In this case the arrangement of the charges $(\wt\alpha_1,\cdots \wt\alpha_{n-2},
\wt\alpha_n)$ follows the corresponding rules for ${\rm N}^{(n-1)}_{\rm higgs}$.
Since $\wt\alpha_{n-1}$ is almost zero we have 
$\beta_{n-1}\simeq \beta_{n-2}$, and thus they must be on the
same side of $\bar\alpha$. 
The condition on 
the charges $\wt\alpha_1,\cdots \wt \alpha_{n-2}, \wt\alpha_n$
for which ${\rm N}^{(n-1)}_{\rm higgs}$ is non-zero can now be written as
\ben \label{eallowednm2}
&&\beta_{k} \cases{\hbox{$< \bar\alpha$ for 
$\wt\alpha_{k+1}>\wt\alpha_{k}$}\cr
\hbox{$>\bar\alpha$ for $\wt\alpha_{k+1}<\wt\alpha_{k}$},
}\, , \qquad \hbox{for $1\le k\le n-3$}\, , \nn
&& \beta_{n-2}, \beta_{n-1}\cases{\hbox{$< \bar\alpha$ for 
$\wt\alpha_{n}>\wt\alpha_{n-2}$}\cr
\hbox{$>\bar\alpha$ for $\wt\alpha_{n}<\wt\alpha_{n-2}$}}\, .
\een
By examining the behavior of $W$ as $ x_{n-1}\to  x_{n-2},  x_n$
we get
\ben \label{ewesee1}
W &\to& -
\langle \wt\alpha_{n-2}, \wt\alpha_{n-1}\rangle \ln| x_{n-2}- x_{n-1}|
\quad \hbox{as $ x_{n-1}\to  x_{n}$}\nn
W &\to& -
\langle \wt\alpha_{n-1}, \wt\alpha_{n}\rangle \ln| x_{n-1}- x_n|
\quad \hbox{as $ x_{n-1}\to  x_{n}$}\, .
\een
Thus in order that they have the same sign so that we have an
extremum we need
\be \label{eweneed}
\wt\alpha_{n-2}<\wt\alpha_{n-1}<\wt\alpha_n \quad \hbox{or} \quad
\wt\alpha_{n-2} > \wt\alpha_{n-1} > \wt\alpha_n\, .
\ee
Combining this with \refb{eallowednm1} we arrive at the result:
\ben \label{eallowednm22}
&&\beta_{k} \cases{\hbox{$< \bar\alpha$ for 
$\wt\alpha_{k+1}>\wt\alpha_{k}$}\cr
\hbox{$>\bar\alpha$ for $\wt\alpha_{k+1}<\wt\alpha_{k}$},
}\, , \qquad \hbox{for $1\le k\le n-1$}\, , \nn
&& \beta_{n-2}, \beta_{n-1} > \bar\alpha \quad \hbox{or} \quad
\beta_{n-2}, \beta_{n-1} < \bar\alpha\, .
\een
The second condition is of course needed to have an almost zero
$\wt\alpha_{n-1}$ in the first place. 
The first condition is the same as the one for getting a non-vanishing
contribution to ${\rm N}^{(n)}_{\rm higgs}(\{\alpha_i\};\sigma)$.
Thus we see that the requirement
for having non-vanishing ${\rm N}^{(n)}_{\rm coulomb}(\{\alpha_i\};\sigma)$ again reduces to
that of having non-vanishing ${\rm N}^{(n)}_{\rm higgs}(\{\alpha_i\};\sigma)$.

We now need to calculate the sign of the contribution. The centers
$1,\cdots n-2, n$ give  a factor of
$(-1)^{\sum_{k=1}^{n-3} \Theta(\wt\alpha_i, \wt\alpha_{i+1})}
(-1)^{\Theta(\wt\alpha_{n-2}, \wt\alpha_{n})}$. 
Using \refb{eweneed} the last factor may be written as
$(-1)^{\Theta(\wt\alpha_{n-2}, \wt\alpha_{n-1})}$.
By studying the behavior 
of $W$ as $ x_{n-1}\to  x_{n-2},  x_n$ we see that the extra contribution
from the sign of $\p^2 W/\p  x_{n-1}^2$ is positive for
$\wt\alpha_{n-2}<\wt\alpha_{n-1}<\wt\alpha_n$ and negative for
$\wt\alpha_{n-2}>\wt\alpha_{n-1}>\wt\alpha_n$. Thus the extra
contribution can be written as $(-1)^{\Theta(\wt\alpha_{n-1}, \wt\alpha_{n})}$.
Combining these factors we get
\be \label{ewtnmps}
{\rm N}^{(n)}_{\rm coulomb}(\{\alpha_i\};\sigma) =
(-1)^{\sum_{i=1}^{n-1} \Theta(\wt\alpha_i, \wt\alpha_{i+1})}\, ,
\ee
in agreement with the formula \refb{higgsfnewest} for
${\rm N}^{(n)}_{\rm higgs}(\{\alpha_i\};\sigma)$. 
This establishes the equality of ${\rm N}^{(n)}_{\rm higgs}(\{\alpha_i\};\sigma)$ and
${\rm N}^{(n)}_{\rm coulomb}(\{\alpha_i\};\sigma)$.

\bigskip

{\bf Acknowledgement:} I wish to thank J.~Manschot and B.~Pioline for
useful correspondence.
This work was
supported in part by the J. C. Bose fellowship of 
the Department of Science and Technology, India and the 
project 11-R\&D-HRI-5.02-0304.

\appendix

\sectiono{Physical interpretation of $g_{\rm higgs}$} \label{sa}

In this appendix we shall show the equivalence between
\refb{higgsfnewest} and the 
formula given in 
\cite{Manschot:2010qz}.
Let $\{\sigma(1), \sigma(2), \cdots \sigma(n)\}$ denote a permutation
of $\{1,\cdots n\}$. 
Associated with each such
permutation we can associate a unique number $s$ and a set  of 
numbers $m_1,\cdots m_{s-1}$ by imposing the following requirements:
\begin{enumerate}
\item $1< m_1 < m_2< \cdots m_{s-1} \le n$.
\item $\sigma(m) > \sigma(m-1)$ for $m\ne m_1, m_2, \cdots m_{s-1}$.
\item $\sigma(m_a) < \sigma(m_a-1)$ for $1\le a\le s-1$.
\end{enumerate}
Physically this partitions
the ordered set $\{\sigma(1),\sigma(2),\cdots\sigma(n)\}$
into $s$ maximally increasing subsequences: 
the $\sigma(i)$'s increase monotonically with $i$
for $i$ between 
$1$ and $m_1-1$, between $m_a$ and $m_{a+1}-1$ for $1\le a\le (s-2)$, and
between $m_{s-1}$ and $n$, but 
between $m_a-1$ and $m_a$ $\forall a$
the monotone increase is broken. 
The expression for ${\rm N}^{(n)}_{\rm higgs}(\{\alpha_i\};\sigma)$ given in
\refb{higgsfnewest} can now be rewritten as
\be \label{enmps}
{\rm N}^{(n)}_{\rm higgs}(\{\alpha_i\};\sigma) = (-1)^{s-1} 
 \prod_{k=m_1, \cdots m_{s-1}} \Theta\left( \alpha_1+
 \cdots \alpha_n,\sum_{i=k}^n \alpha_{\sigma(i)}\right)
 \prod_{k=2\atop k\ne m_1,\cdots m_{s-1}}^n \Theta\left(
 \sum_{i=k}^n \alpha_{\sigma(i)},\alpha_1+\cdots \alpha_n\right)
 \ee

Let us
define a set of vectors $\beta^{(1)},\cdots \beta^{(s)}$
as follows:
\be \label{edefbeta}
\beta^{(1)} = \sum_{i=m_{s-1}}^n  \alpha_{\sigma(i)}, \quad
\beta^{(a)} = \sum_{i=m_{s-a}}^{m_{s-a+1}-1} \alpha_{\sigma(i)} \quad \hbox{for}
\quad 2\le a\le s-1, \quad \beta^{(s)} = \sum_{i=1}^{m_1-1}
\alpha_{\sigma(i)}\, .
\ee
This allows us
to associate to every permutation $\sigma(i)$ a unique set of vectors 
$\{\beta^{(a)}\}$. 
For example if
for $n=4$ we consider the permutation $(2134)$ then the partition
of $(2134)$ containing maximally increasing
subsequences are $\{\{2\}, \{1,3,4\}\}$, giving $\beta^{(1)}=\alpha_1+
\alpha_3+\alpha_4$,
$\beta^{(2)}=\alpha_2$. 
The product of the $\Theta$'s in \refb{higgsfnewest} restricts the
sum over permutations to
a set $K$ of permutations 
satisfying the following conditions:
\begin{enumerate}
\item The first set of $\Theta$ functions in \refb{enmps}
ensure that the
vectors $\{\beta^{(a)}\}$ associated with the permutation $\sigma$ 
should satisfy
\be \label{eimpeq}
\left\langle \sum_{a=1}^b \, \beta^{(a)}, 
\alpha_1+\cdots + \alpha_n\right\rangle > 0 \quad
\forall \quad b \ \hbox{with} \ 1\le b\le s-1\, .
\ee
\item We can associate with the permutation $\sigma$ many other 
partitions containing increasing
subsequences which are not maximal, by dropping the third condition
$\sigma(m_a) < \sigma(m_a-1)$. Thus for example for the permutation
(2134) discussed above, examples of partitions containing
non-maximal increasing subsequences
are $\{\{2\}, \{1\}, \{3\}, \{4\}\}$, $\{\{2\}, \{1,3\}, \{4\}\}$ and
$\{\{2\}, \{1\}, \{3,4\}\}$. The second set of $\Theta$-functions in \refb{enmps}
guarantee
that if we construct the $\beta^{(a)}$'s for any such
partition following the same procedure, then the condition
\refb{eimpeq} must fail for at least one $b$. 
\end{enumerate}
Once we have identified the set $K$ of permutations satisfying these
properties, \refb{higgsfnewest} reduces to:
\begin{equation} 
\label{higgsfnew}
 g_{\rm higgs}(\alpha_1,\dots ,\alpha_n)=(-y)^{-1+n
 -\sum_{i<j} \alpha_{ij}} \,
 (y^2-1)^{1-n} \, 
 \sum_{\sigma\in K}(-1)^{s-1} (-y)^{
 2\sum_{l<k\atop\sigma(l)<\sigma(k)}
 \alpha_{\sigma(l)
 \sigma(k) }}\, .
\end{equation}
This is the formula for $g_{\rm higgs}$ derived in \S3.3 of
\cite{Manschot:2010qz}.\footnote{In the 
last term \cite{Manschot:2010qz} had $y^{\cdots}$ instead of
$(-y)^{\cdots}$. 
For physical $\alpha_{ij}$'s which are integers
the two formul\ae\ give identical results since the exponent is an even integer.}
The original Reineke formula\cite{MR1974891} 
corresponds to summing over 
many more terms corresponding to all increasing sequences (\i.e.\
not just the maximal increasing sequences), but
it was shown in \cite{Manschot:2010qz} that
the contribution from
many of
these terms cancel and at the end only the contribution from the terms
given in \refb{higgsfnew}, corresponding to maximal increasing 
sequences, survive.

\sectiono{Proof of the $\Theta$ identity} \label{sb}

In this appendix we shall prove the identity \refb{eid1}. Let us denote
the left hand side of \refb{eid1} by $P(N,c)$, \i.e.\
\ben \label{epnc}
&& P(N, c) \nn
&\equiv & \prod_{i=1}^{N-1} \bigg[ \Theta\Big( f (i+1)- c
\Big) \Theta\Big( f (i+1)- f (i)\Big) 
- \Theta\Big(c- f (i+1)\Big) \Theta
\Big( f (i) - f (i+1)\Big)\bigg]\nn
&& +\sum_{k=1}^{N-1} \sum_{I_1=2}^{N}
\sum_{I_2=I_1+1}^{N} \sum_{I_3=I_2+1}^{N} \cdots
\sum_{I_{k}=I_{k-1}+1}^{N} \Bigg\{\Theta(c- f (I_1)) \prod_{\ell=1}^{k-1} 
\Theta\Big( f (I_\ell) -  f (I_{\ell+1})\Big)\nn
&&
\prod_{i=1}^{I_{1}-2}
\bigg[ \Theta\Big( f (i+1)
-c\Big) \Theta\Big( f (i+1)- f (i)\Big) 
- \Theta\Big(c -  f (i+1)\Big) \Theta\Big( f (i) - f (i+1)\Big)\bigg]
\nn
&&\prod_{\ell=1}^{k-1} 
\prod_{i=I_\ell}^{I_{\ell+1}-2}
\bigg[ \Theta\Big( f (i+1)-  f (I_\ell)
\Big) \Theta\Big( f (i+1)- f (i)\Big) 
- \Theta\Big( f (I_\ell)- f (i+1)\Big) \Theta
\Big( f (i) - f (i+1)\Big)\bigg]\nn
&& \prod_{i=I_k}^{N-1} \bigg[ \Theta\Big( f (i+1)-  f (I_k)
\Big) \Theta\Big( f (i+1)- f (i)\Big) 
- \Theta\Big( f (I_k)- f (i+1)\Big) \Theta
\Big( f (i) - f (i+1)\Big)\bigg]
\Bigg\} \, .\nn
\een
We shall assume that \refb{eid1} is valid up to a certain value of
$N$, e.g. we have
\be \label{ewehave}
P(M, c) =\prod_{i=1}^{M-1} \Theta( f (i+1)- f (i))\, ,
\quad \hbox{for $M\le N-1$}
\ee 
and then show that \refb{ewehave} also holds for $M=N$. Now 
in  the $k\ge 1$ terms in \refb{epnc}  the
sum over $k$ and $I_2,\cdots I_k$ for fixed $I_1$, after factoring out
the $I_1$ dependent but $k$ independent terms, has the same structure
as $P(N-I_1+1, f(I_1))$ with $ f (i)\to  f (i+I_1-1)$. This gives
\ben \label{epnc1a}
&& P(N, c)\nn
&=& \prod_{i=1}^{N-1} \bigg[ \Theta\Big( f (i+1)- c
\Big) \Theta\Big( f (i+1)- f (i)\Big) 
- \Theta\Big(c- f (i+1)\Big) \Theta
\Big( f (i) - f (i+1)\Big)\bigg]\nn
&& + \sum_{I_1=2}^{N} \Theta(c- f (I_1)) 
P(N-I_1+1, f(I_1))\Big|_{f(i)\to f(i+I_1-1)}\nn
&& 
\prod_{i=1}^{I_{1}-2}
\bigg[ \Theta\Big( f (i+1)
-c\Big) \Theta\Big( f (i+1)- f (i)\Big) 
- \Theta\Big(c -  f (i+1)\Big) \Theta\Big( f (i) - f (i+1)\Big)\bigg]  \, .\nn
\een
Using
\refb{ewehave} the result can be
expressed as
\ben\label{epnc1}
&&P(N, c)\nn
&=& \prod_{i=1}^{N-1} \bigg[ \Theta\Big( f (i+1)- c
\Big) \Theta\Big( f (i+1)- f (i)\Big) 
- \Theta\Big(c- f (i+1)\Big) \Theta
\Big( f (i) - f (i+1)\Big)\bigg] \nn &&
+ \sum_{I_1=2}^{N} \Theta(c- f (I_1)) 
\prod_{i=I_1}^{N-1} \Theta( f (i+1)- f (i))\nn
&& 
\prod_{i=1}^{I_{1}-2}
\bigg[ \Theta\Big( f (i+1)
-c\Big) \Theta\Big( f (i+1)- f (i)\Big) 
- \Theta\Big(c -  f (i+1)\Big) \Theta\Big( f (i) - f (i+1)\Big)\bigg]
\, . \nn
\een
Using the relation
\be \label{euse1}
\Theta(x)\Theta(y)-\Theta(-x)\Theta(-y)
= \Theta(x)(1-\Theta(-y))- (1-\Theta(x))\Theta(-y) = \Theta(x)-\Theta(-y)\, ,
\ee
we can simplify \refb{epnc1} to
\ben \label{eed2}
&& P(N,c) \nn
&=& \prod_{i=1}^{N-1} \left[\Theta\Big( f (i+1)- f (i)\Big) - 
\Theta\Big(c - f (i+1)\Big)\right] \nn
&& +\sum_{I_1=2}^{N} \Theta\Big(c- f (I_1)\Big) 
\prod_{i=I_1}^{N-1} \Theta\Big( f (i+1)- f (i)\Big)
\prod_{i=1}^{I_1-2}
\left[\Theta\Big( f (i+1)- f (i)\Big) - \Theta\Big(c - f (i+1)
\Big)\right]\, . \nn
\een
We can manipulate this by separating out the $I_1=N$ term in the sum
and combining it with the first term. This gives
\ben \label{eed3}
&& P(N,c) \nn
&=& \prod_{i=1}^{N-2} \left[\Theta\Big( f (i+1)- f (i)\Big) - 
\Theta\Big(c - f (i+1)\Big)\right]  \nn &&
\left[ \Theta\Big( f (N)- f (N-1)\Big) - 
\Theta\Big(c - f (N)\Big) + \Theta\Big(c- f (N)\Big)
\right]
\nn
&& +\sum_{I_1=2}^{N-1} \Theta\Big(c- f (I_1)\Big) 
\prod_{i=I_1}^{N-1} \Theta\Big( f (i+1)- f (i)\Big)
\prod_{i=1}^{I_1-2}
\left[\Theta\Big( f (i+1)- f (i)\Big) - \Theta\Big(c - f (i+1)
\Big)\right] \nn
&=& 
\Theta\Big( f (N)- f (N-1)\Big) \Bigg\{
\prod_{i=1}^{N-2} \left[\Theta\Big( f (i+1)- f (i)\Big) - 
\Theta\Big(c - f (i+1)\Big)\right] \nn &&
+ \sum_{I_1=2}^{N-1} \Theta\Big(c- f (I_1)\Big) 
\prod_{i=I_1}^{N-2} \Theta\Big( f (i+1)- f (i)\Big)
\prod_{i=1}^{I_1-2}
\left[\Theta\Big( f (i+1)- f (i)\Big) - \Theta\Big(c - f (i+1)
\Big)\right] 
\Bigg\}\, .\nn
\een
We now notice that the term inside the $\{~\}$ has the same form as the
right hand side of \refb{eed2} with $N$ replaced by $N-1$. Thus we
can manipulate it again in the same way, pulling out a factor
of $\Theta\Big( f (N-1)- f (N-2)\Big)$ and replaing $N$ by $N-1$
again in the remaining factor. Repeating this process we arrive
at the result:
\be \label{eed4}
P(N,c) =\prod_{i=1}^{N-1} \Theta\Big( f (i+1)- f (i)\Big)\, ,
\ee
which is the desired result.

\end{document}